%% file: main.tex
\documentclass[12pt,letterpaper]{lsuetd}
\usepackage{setspace,graphics,dsfont,verbatim,paralist,indentfirst}
\usepackage{graphicx}
\usepackage{amssymb}
\usepackage{comment}
\usepackage{amsmath}
\usepackage{csquotes}
\usepackage{titlesec}

\usepackage{epstopdf}
\usepackage{physics}
\usepackage{subfigure}
\usepackage{xcolor}
\usepackage{hyperref}
\hypersetup{
    colorlinks=true,
    citecolor=red,
    filecolor=black,
    linkcolor=blue,
    urlcolor=black,
    linktoc=all
}
\titleformat*{\section}{\normalsize\bfseries}
\titleformat*{\subsection}{\normalsize\bfseries}
\makeatletter

\begin{document}
\renewcommand\@pnumwidth{1.55em}
\renewcommand\@tocrmarg{9.55em}
\renewcommand*\l@chapter{\@dottedtocline{0}{1.5em}{2.3em}}
\renewcommand*\l@figure{\@dottedtocline{1}{0em}{3.1em}}
\let\l@table\l@figure
\renewcommand{\bibname}{References}
\pagenumbering{roman}
\thispagestyle{empty}
\begin{center}
%The title page is first created.
\textbf{\fontsize{14}{12} \selectfont THIRD GENERATION GAMMA CAMERA SPECT SYSTEM}

\vfill
\doublespacing
A Thesis\\
\singlespacing
Submitted to the Graduate Faculty of the\\
Louisiana State University and\\
Agricultural and Mechanical College\\
in partial fulfillment of the\\
requirements for the degree of\\
Master of Science in Medical Physics and Health Physics\\
\doublespacing
in\\
                                       
The Department of Physics \& Astronomy\\
\singlespacing
\vfill

by\\
Narayan Bhusal\\
B.Sc. \& M.Sc., Tribhuvan University, Nepal, 2013\\
%If necessary, copy and paste the previous line here to include a master's degree.
December 2018
\end{center}
\pagebreak

\singlespacing
\tableofcontents
\pagebreak
%The code below generates the List of Tables and adds it to the Table of Contents.
\renewcommand\@pnumwidth{1.55em}
\renewcommand\@tocrmarg{8.55em}
%\addcontentsline{toc}{chapter}{\hspace{-1.5em} LIST OF TABLES \vspace{12pt}}

%\pagebreak
%The code below generates the List of Figures and adds it to the Table of Contents.
%\addcontentsline{toc}{chapter}{\hspace{-1.5em} LIST OF FIGURES \vspace{12pt}}
%\listoffigures
%\pagebreak
%The List of Nomenclature may be included here, if desired.

%\chapter*{List of Nomenclature}
%\doublespacing
%\vspace{0.55ex}
%Provide the definitions of the symbols used in your thesis or dissertation here.
%\addcontentsline{toc}{chapter}{\hspace{-1.5em} LIST OF NOMENCLATURE \vspace{12pt}}
%\pagebreak

%The code below adds the Abstract and places it within the Table of Contents.
\renewenvironment{abstract}{{\hspace{-2.2em} \huge \textbf{\abstractname}} \par}{\pagebreak}
\addcontentsline{toc}{chapter}{\hspace{-1.5em} ABSTRACT}
\begin{abstract}
\vspace{0.55ex}
\doublespacing
Single Photon Emission Computed Tomography (SPECT) is a non-invasive imaging modality, frequently used in myocardial perfusion imaging. The biggest challenges facing the majority of clinical SPECT systems are low sensitivity, poor resolution, and the relatively high radiation dose to the patient. New generation systems (GE Discovery, DSPECT) dedicated to cardiac imaging improve sensitivity by a factor of 5-8. The purpose of this work is to investigate a new gamma camera design with 21 hemi-ellipsoid detectors each with a pinhole collimator for Cardiac SPECT for further improvement in sensitivity, resolution, imaging time, and radiation dose.

To evaluate the resolution of our hemi-ellipsoid system, GATE Monte-Carlo simulations were performed on point-sources, rod-sources, and NCAT phantoms. The purpose of the point-source simulation is to obtain operating pinhole diameter by comparing the average FWHM (Full-width-half-maximum) of the flat-detector system with a curved hemi-ellipsoid detector system. The operating pinhole diameter for the curved hemi-ellipsoid detector was found to be 8.68mm.  System resolution is evaluated using reconstructed rod-sources equally spaced within the region of interest. The results were compared with results of the GE discovery system available in the literature. The system performance was also evaluated using the mathematical anthropomorphic NCAT (NURBS-based Cardiac Torso) phantom with a full (clinical) dose acquisition (25mCi) for 2 mins and an ultra-low-dose acquisition of 3mCi for 5.44mins.

On rod-sources, the average resolution after reconstruction with resolution recovery in the entire region of interest (ROI) for cardiac imaging was 4.44mm, with a standard deviation of 2.84mm, compared to 6.9mm reported for GE Discovery (Kennedy et al., JNC, 2014). For NCAT studies improved sensitivity allowed a full-dose (25mCi) 2 min acquisition (ELL8.68mmFD) which yielded 3.79M LV counts. This is ~3.35 times higher compared to 1.13M LV counts acquired in 2 mins for clinical full-dose for state-of-the-art DSPECT. The increased sensitivity also allowed an ultra-low dose acquisition protocol (ELL8.68mmULD). This ultra-low-dose protocol yielded ~1.23M LV-counts which was comparable to the full-dose 2min acquisition for DSPECT. The estimated NCAT average FWHM at the LV wall after 12 iterations of the OSEM reconstruction was 4.95mm and 5.66mm around the mid-short-axis slices for ELL8.68mmFD and ELL8.68mmULD respectively.

\end{abstract}

\pagenumbering{arabic}
\addtocontents{toc}{\vspace{12pt} \hspace{-1.8em} CHAPTER \vspace{-1em}}
\singlespacing
\setlength{\textfloatsep}{12pt plus 2pt minus 2pt}
\setlength{\intextsep}{6pt plus 2pt minus 2pt}

\chapter{Introduction}
\doublespacing
\input{chapter1}

\pagebreak
\singlespacing
\chapter{Methods and Results}
\doublespacing
\input{chapter2}

\pagebreak
\singlespacing
\chapter{Conclusion and Future Works}
\doublespacing
\input{chapter3}

\pagebreak
\singlespacing
%\chapter{Chapter 5 Title}
%\doublespacing
%\input{chapter5}
%\pagebreak
%\singlespacing
%\chapter{Conclusion}
%\doublespacing
%\input{chapter6}
%\pagebreak
%\singlespacing
%To insert additional chapters, copy the previous five lines, using chapterX as the argument of the 
%\input command for Chapter X, where X=6,7,8,...
\addtocontents{toc}{\vspace{12pt}}
\addcontentsline{toc}{chapter}{\hspace{-1.6em} REFERENCES}
\bibliographystyle{unsrt}
\bibliography{main.bib}

% \pagebreak
% \singlespacing
% \addtocontents{toc}{\vspace{12pt} \hspace{-1.8em} APPENDIX \vspace{-1em}}
% \appendix
% \chapter{Author's Publications}
% \vspace{0.5em}
% \input{appendix_publicat}
% \chapter{Reuse Permission}
% \vspace{0.5em}
% \input{appendix_reuse}
%If you need to insert additional appendices, copy the previous four lines, using appendixY as the
%argument of the \input commnd for Appendix Y, for Y=C,D,E,...
%Finally, the vita section is created and included in the Table of Contents.
% \chapter*{Vita}
% \doublespacing
% \setlength{\parindent}{1.75em}
% \vspace{0.2em}
% \addtocontents{toc}{\vspace{12pt}}
% \addcontentsline{toc}{chapter}{\hspace{-1.5em} VITA}
% David was born and raised in a small town in Louisiana. Dr. Smith will be working as a.

\end{document}

%% file: chapter1.tex
\label{ch1}
Medical Imaging is a process of shedding light on the biological tissues which are otherwise invisible to naked eyes. There have been rapid advances in the field of medical imaging since German scientist William Conrad Röntgen produced the first X-ray image (an x-ray radiograph of his wife’s hand) in 1895 \cite{bushberg2011essential}. Radiography, computed tomography (CT), magnetic resonance imaging (MRI), single-photon emission computed tomography (SPECT), and positron emission tomography (PET) are the major imaging modalities that are being used frequently as diagnostic tools in the field of medicine. CT and MRI are anatomical imaging modalities, meaning they render the anatomical details in the images. However, SPECT and PET are the functional imaging systems that are aimed at exploring the physiology of the tissue of interest. Usually, SPECT and PET are combined with an anatomical imaging modality to put together the functional/physiological information and anatomical details. Combining anatomy with physiology is very important from the diagnostic point of view \cite{wernick2004emission}. SPECT and PET use radioactive material as the tracers to produce the map of concentration of radionuclide uptake, which is why they are categorized as nuclear imaging modalities. For example, in myocardial perfusion imaging (MPI), the concentration of radiopharmaceutical tells us how well the blood is flowing through the myocardium (heart muscle) which is indeed functional/physiological information.

\section{Background of Gamma Camera}
\label{sec11}
\begin{figure}[ht!]
    \centering
    \includegraphics[width=0.96\textwidth]{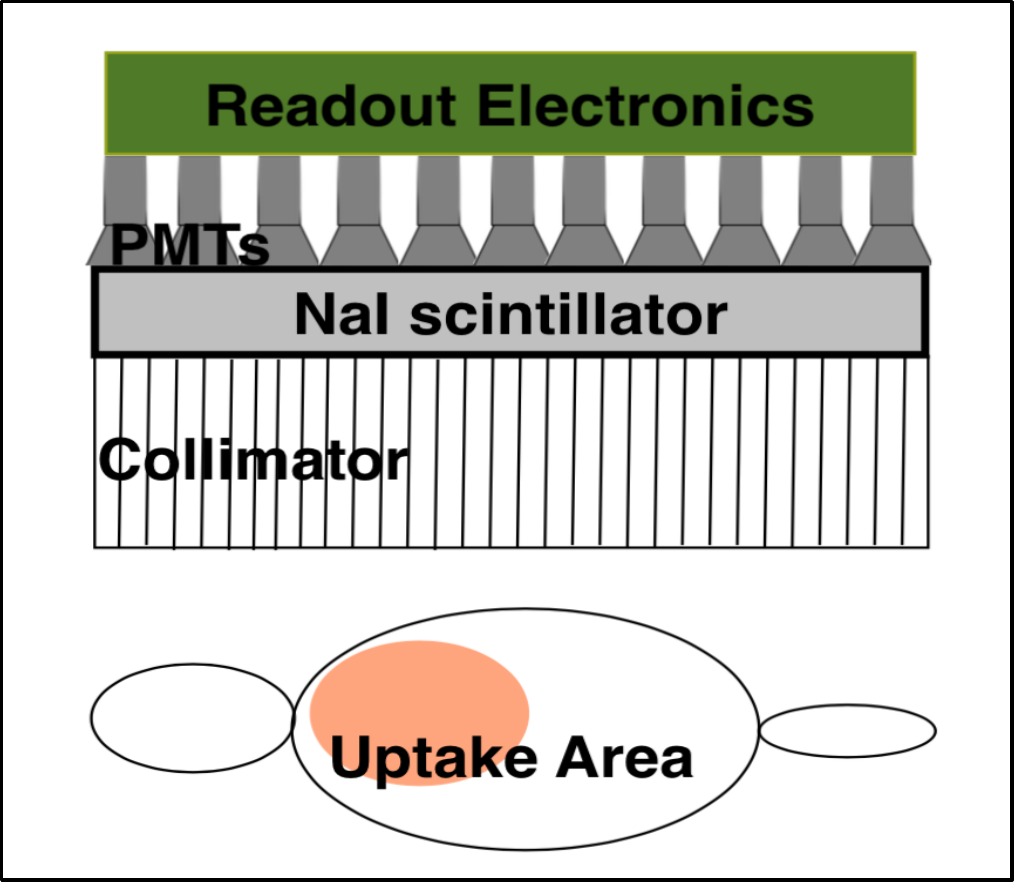}
    \caption[Schematic diagram of Anger gamma camera.]{Schematic diagram of Anger gamma camera. Collimator, scintillator, light guide, PMTs and readout electronics are displayed. The gamma photons emitted from the uptake area are registered by the readout electronics.}
    \label{ch1_gammaCam}
\end{figure}

A Gamma camera is a device that is used to produce the distribution of radioisotope within the tissue of interest by detecting gamma photons emitted during radioactive decay. The first gamma camera was proposed by Hal Anger (University of California, Berkeley) in 1958 which uses lead collimators, thallium-activated sodium iodide crystal as scintillator, photomultiplier tubes (PMT), and electronic readout mechanism \cite{wernick2004emission}. Different variants of this camera are still in use in a large number of clinics across the globe which are called Anger Camera. The schematic diagram of the Anger camera is shown in Figure \ref{ch1_gammaCam}. The basic working principle of a gamma camera is described as follows: The radionuclides absorbed by the tissue in the uptake area emit gamma photons (which are called singles) as they undergo radioactive decay. The photons incident on the collimator within the acceptance angle hit the scintillator crystal. The visible light photons are generated as the gamma photons are absorbed by scintillator material which is fed to photo-multiplier tubes (PMTs) through a light guide. A readout electronics connected to PMTs generates a 2-dimensional intensity distribution which is called projection image.

Gamma camera is either used for planar imaging or tomographic imaging. Planar gamma imaging is mainly used for thyroid imaging, bone scan, and ventilation/perfusion imaging of lung, etc. Myocardial perfusion imaging (MPI), cerebral blood flow imaging, and tumor imaging, etc. are the main applications of tomographic gamma imaging described below.

\section{SPECT Imaging}
\label{sec12}
As the name suggests, Single Photon Emission Computed Tomography or SPECT combines basic elements of gamma camera (Anger Camera) with the tomographic image reconstruction. Tomography is a process of combining 2-dimensional projection images acquired at different angles to render the 3-dimensional information. The beauty of tomographic imaging is that you get volumetric information of tissue being imaged non-invasively which is a valuable asset from the diagnostic point of view.

\begin{figure}[ht!]
    \centering
    \includegraphics[width=0.75\textwidth]{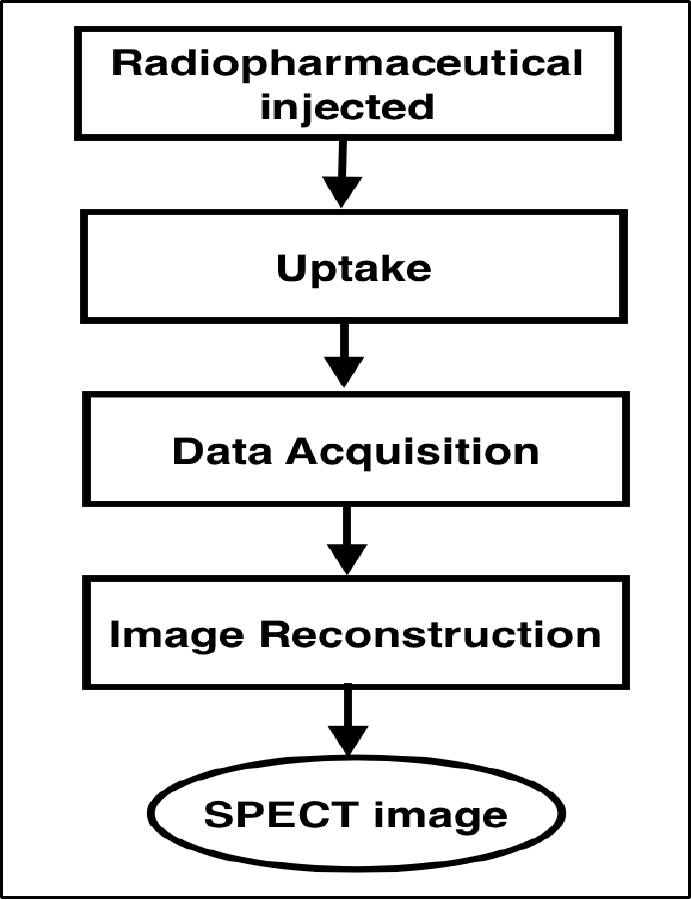}
    \caption[Schematic diagram of workflow of a SPECT imaging.]{Schematic diagram of workflow of a SPECT imaging.}
    \label{ch1_wf}
\end{figure}

A typical workflow of a SPECT imaging is shown in Figure \ref{ch1_wf}. First of all, radiopharmaceutical is injected into the blood stream of the patient. Radiopharmaceutical is the combination of radionuclide and the biological molecule. The function of biological molecule is to transport the attached radionuclide to the tissue/organ of interest. 99mTc-MDP (Methyl Diphosphonate) is used as a radiotracer for bone imaging. Whereas, 99mTc-Sestamibi and 99mTc-ECD (Ethyl Cysteinate) are the commonly used radiopharmaceuticals for MPI and brain imaging respectively. Ga-67 and In-111 based radiopharmaceuticals are used for tumor imaging \cite{wernick2004emission}.

A period of time needed for the radioisotopes to accumulate in the region of interest (ROI) is called uptake time. In some cases, patient is exercised during this period depending upon the type of the study. Typical uptake time varies from 20 minutes to 30 minutes. For MPI, a common cardiac scan using SPECT, standard uptake time is 20 minutes \cite{kincl2015cadmium}.

After the radiation uptake, patient is then taken to the gamma camera and several projection images are acquired at various angles around the region of interest (ROI). The projection angles are chosen such that every part of the ROI is properly covered. Often times these acquired projections need some kind of post-processing in order to meet the requirements of image reconstruction algorithm. 

The projections thus obtained are the inputs to the image reconstruction algorithm. There are two major categories of reconstruction algorithms: analytical reconstruction, and iterative image reconstructions \cite{zeng2001image}. We use MLEM (Maximum Likelihood Expectation Maximization), OSEM (Ordered Subset Expectation Maximization) reconstruction algorithm which are the examples of iterative image reconstruction. These two reconstruction algorithms are discussed briefly in Section \ref{sec14} of this chapter. Further details about the algorithms can be found in \cite{dempster1977maximum, lange1984reconstruction, hudson1994accelerated}.

SPECT is mainly used in myocardial perfusion imaging (MPI) whose function is to test how well the blood is flowing through the heart muscle (called myocardium). The cold spots in the image (where the density of radioactive tracer is less) indicate the lack of sufficient blood flow in the area. Doctors use this information to assess myocardial infarction, and coronary artery disease (CAD). Two kinds of MPI tests are performed routinely: rest study and stress study. The projections are acquired when the person is resting for the rest study. Whereas the stress study is performed when the patient is either physically or pharmacologically exercised \cite{jhm2018}. The standard radiation doses injected to the patient for rest and stress study are 15mCi and 25mCi respectively \cite{kalluri2015multi}.

\section{SPECT Hardware}
\label{sec13}
The most important hardware of SPECT system include collimator, scintillator, light guide, and electronic readout mechanism. The performance of imaging system depends heavily on these devices. Dedicated cardiac SPECT system has gone through a lot of hardware changes in past few decades which will be discussed in Section \ref{sec16}.

Since the emission of photons from the tissue that has taken up radioactive tracer is isotropic in nature, collimators are used to select or reject the photons emitted based on the direction. Only the photons travelling in a direction within the acceptance angle of the collimators are accepted which is the fundamental principle of collimation. There are several types of collimators: parallel hole collimator, pinhole collimator, diverging collimator, converging collimator, slanthole collimator, and fan-beam collimator \cite{smith2010introduction}. The most commonly used collimator is the parallel hole collimator. Collimators are usually made of high density, and high atomic number (Z) materials like lead, tungsten etc. The biggest drawback of having to use the collimators is that a very large proportion of gamma photons will be lost. That’s why the SPECT system has very low sensitivity. We could, however, increase the size of the collimator holes to improve sensitivity. But the spatial resolution, on the other hand, becomes poorer as the collimator is opened up more.  
	
The function of scintillator crystal is to absorb the gamma photon incident on it and generate a visible light photon. The phenomenon is called scintillation. The quality of scintillator crystal is assessed based on how well it can stop the gamma photon and allow the passage of secondary light photon. Ideal crystal absorbs every gamma photon successfully and is completely transparent to the visible light photon \cite{wernick2004emission}. Some of the most efficient scintillator materials used in SPECT imaging are NaI, CsI etc. We use CsI as the scintillator crystal for this project.
	
Light guide is another important component of SPECT hardware. Its function is to send visible light photons generated in scintillator to the detectors like PMTs, SiPMs etc. Light guide is made of material which is transparent/non-absorbing to the visible light. Detectors in combination with the attached electronic readout mechanism produces a 2-dimensional intensity pattern called a projection. A series of such projections are acquired before computing the reconstructed volumetric image.

\section{Image Reconstruction}
\label{sec14}
Image reconstruction is the process of combining projection images to produce a meaningful 3D/volumetric image of ROI. There are two categories of image reconstruction algorithms: Analytical reconstruction and Iterative reconstruction. Filtered back projection, Inverse Radon Transform, and Direct Fourier methods are some examples of analytical reconstruction algorithms \cite{wernick2004emission}. Maximum Likelihood Expectation Maximization (MLEM), and Ordered Subset Expectation Maximization (OSEM) are the two iterative reconstruction algorithms we used in this project. 
	
Iterative image reconstruction begins with some initial image or initial intensity distribution. Assuming that an initial estimate of the 3D volume to be reconstructed, different projections are generated by using a standard algorithm called forward-model. In this forward-model, different physical effects such as collimator resolution compensation, and tissue attenuation correction etc. can be incorporated. Then the estimated projections are compared with true input projections (measured projections) supplied to the reconstruction algorithm. Projection space error thus generated is transformed or projected to reconstruction space which is used to update the initial value image. This is done iteratively by assuming the last updated image as the initial value image until terminated by user or desired convergence \cite{zeng2001image}. MLEM is the simplest form of iterative reconstruction algorithm. This algorithm was first proposed in 1977 by Dempster et. al \cite{dempster1977maximum}.  Lange and Carson in 1984 demonstrated first the use of MLEM in emission tomography (ET) \cite{lange1984reconstruction}. The major problem with MLEM is slow convergence and it is computationally expensive. OSEM is a block iterative reconstruction algorithm in which projections are grouped into mutually exclusive subsets. Reconstruction algorithm is applied to each of the subsets in a sequence. OSEM algorithm was developed by Hudson and Larkin in 1994 \cite{hudson1994accelerated}. It is a modified form of MLEM algorithm which converges much faster than MLEM if the subsets are chosen effectively. If all subsets of OSEM are combined to a single subset, the algorithm reduced to MLEM. Problem with OSEM is that the performance of this algorithm depends on how effectively the subsets are created \cite{wernick2004emission}.
	
For this thesis we are using pre-existing multi-pinhole MLEM/OSEM reconstruction code which compensates for attenuation and collimator resolution written by Dr. Dey and used by other groups \cite{kalluri2015multi, dey2011comparing, chan2016impact}.

\section{Performance Measures}
\label{sec15}
\begin{figure}[ht!]
    \centering
    \includegraphics[width=0.75\textwidth]{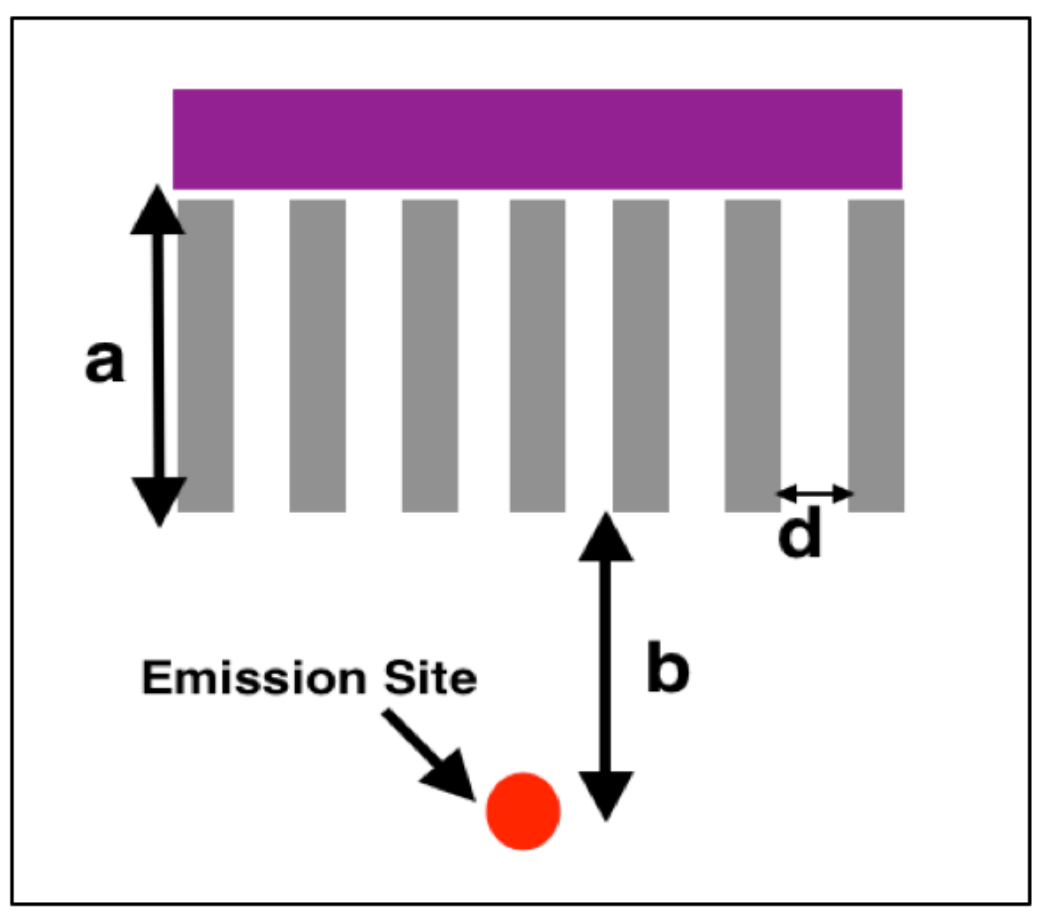}
    \caption[Schematic diagram of gamma camera with parallel hole collimation.]{Schematic diagram of gamma camera with parallel hole collimation.}
    \label{ch1_phc}
\end{figure}

Spatial resolution and sensitivity are the most important performance measures for SPECT system.  Spatial resolution is the quantitative measure of the smallest resolvable physical distance in the image.  The standard way of quantifying spatial resolution is to estimate the full width half maximum (FWHM) at the point of interest. FWHM, in fact, is estimated by analyzing the point spread function at a point inside the region of interest (ROI). System spatial resolution of a gamma camera consists of two components: collimator resolution, and intrinsic detector resolution. The quadrature sum of two resolutions gives the system resolution \cite{cherry2012physics}.  A brief discussion about spatial resolution and photon detection sensitivity of SPECT system using parallel hole collimation (as shown in Figure \ref{ch1_phc}) is presented below.

The collimator resolution of above system is given by,
\begin{equation}
    \operatorname{Res}_{\text {coll }}=d\left(1+\frac{b}{a}\right)
\end{equation}
Let $R_I$ be the intrinsic detector resolution of above system. Intrinsic detector resolution is the measure of how well you can localize the scintillation event in the detector using the available readout mechanism. Ideally it should contain information like depth of interaction. The factor $(b/a)$ in these equations is the inverse of magnification.
\begin{equation}
    \operatorname{Res}=\sqrt{d^{2}\left(1+\frac{b}{a}\right)^{2}+R_{I}^{2}}
\end{equation}
Sensitivity is the measure of relative number of emitted photons reaching the detector. This quantity ultimately determines how accurately the true-positives and false-negatives are predicted using the imaging system. Mathematically it can be expressed as \cite{peterson2012advances},
\begin{equation}
    S \propto \frac{d^{2}}{b^{2}}
\end{equation}

Since image reconstruction algorithm is an integral part of SPECT imaging system, the quality of it is also an important measure of performance. Ideally you would want the reconstructed image to be identical to true or expected image. However, the radioactive emission is random in nature which is why a statistical analysis of performance is needed. Most commonly, bias and variance are estimated. Bias is the measure of deviation of reconstructed mean from the true mean. And, variance measures the distribution or spread of the estimate. In ideal case of both bias and variance, you would estimate the mean reconstructed image with 100\% accuracy every time with independently acquired projections which is in reality impossible. So, it is always important have a balance between two which is known as bias-variance-tradeoff. Bias-variance analysis specific to the system proposed and analyzed in this thesis is presented in Section \ref{sec23}.

\section{Advances in Dedicated Cardiac SPECT}
\label{sec16}
Dedicated cardiac SPECT systems have gone through rapid changes and advancement in last couple of decades. The changes in hardware and software aspects are quite significant that resulted in reduction of acquisition time from $\sim15$ minutes to $\sim2$ minutes \cite{slomka2009advances}. Some of the key hardware changes were made in collimator design. The conventional collimators are now replaced by multi-pinhole design which improved the sensitivity of SPECT system by many fold. Other important changes made to the system include scanner geometry, scintillator crystal which led to improved sensitivity of photon detection \cite{slomka2012advances}.
	
The SPECT systems using standard Anger-camera features are called Generation-I systems. As discussed earlier, Gen-I cameras have very low sensitivity and poor resolution. System geometry of first generation gamma camera is already discussed in Section \ref{sec11}. These cameras use two-day protocol for a cardiac scan in which 25mCi stress study is followed by 15mCi rest study. Long acquisition times, low sensitivity and spatial resolution, and high radiation dose to the patient are major issues with the first generation gamma cameras \cite{slomka2009advances}.

\begin{figure}[ht!]
    \centering
    \includegraphics[width=0.75\textwidth]{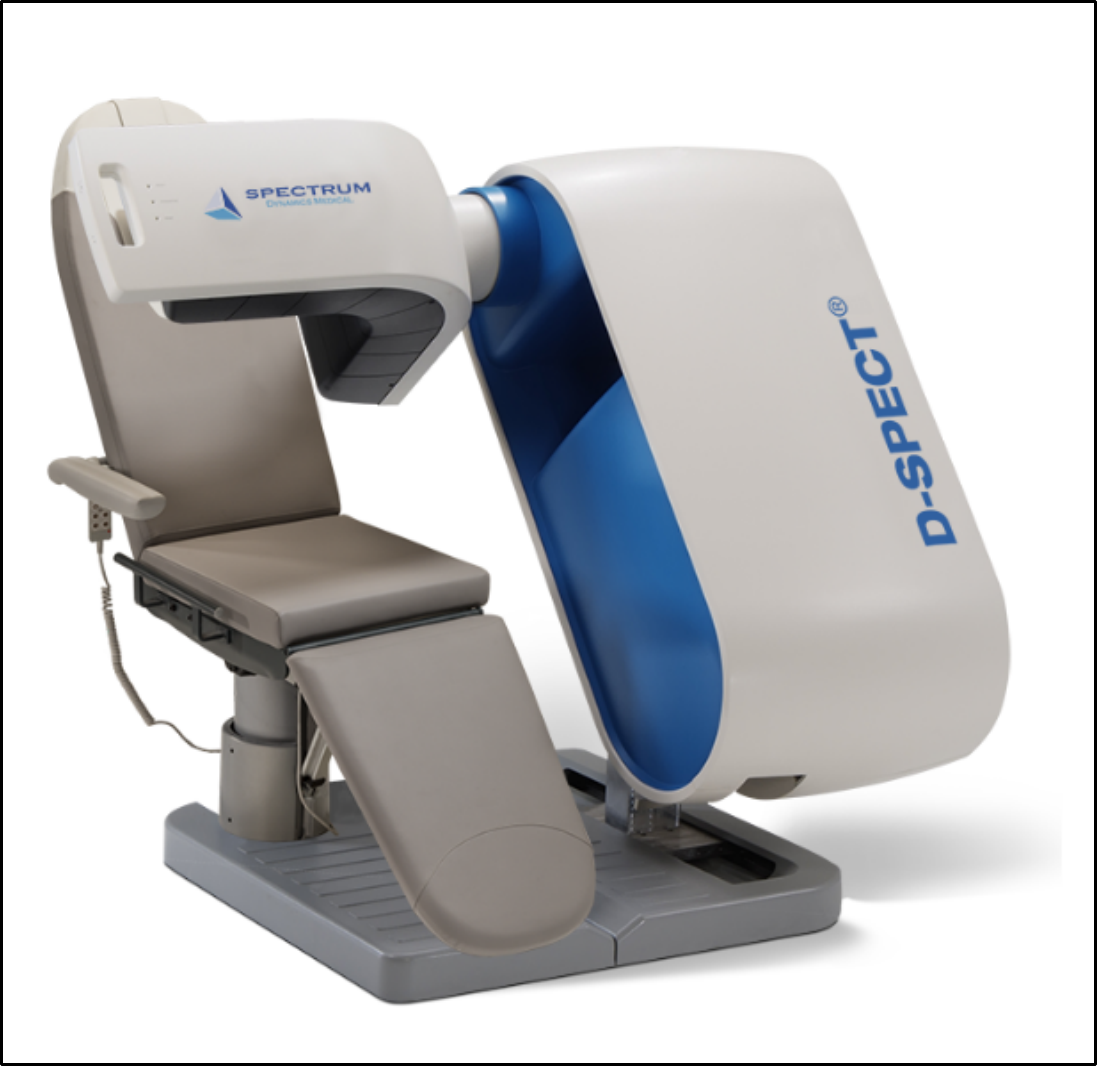}
    \caption[DSPECT, dedicated cardiac SPECT system. This uses 9 parallel hole collimators arced around heart18.]{DSPECT, dedicated cardiac SPECT system. This uses 9 parallel hole collimators arced around the heart \cite{spectrum2018}}
    \label{ch1_dspect}
\end{figure}

A great deal of improvement in system performance is seen since the advent of Generation-II systems. DSPECT (Dynamic SPECT) is a dedicated cardiac SPECT system that uses parallel hole collimator design which is shown in Figure \ref{ch1_dspect}. It is a Generation-II system which has 5-8 times better sensitivity than the first generation Anger-camera \cite{smith2013recent}. DSPECT is a gantry static geometry in which collimators move during data acquisition. It uses CZT (Cadmium Zinc Telluride) detectors (higher detection efficiency), and tungsten collimators. It is a single shot measurement in which all nine projections are acquired at one-go. One of the other key features of DSPECT is comfortable patient positioning, reduced distance between detector and body, reduced radiation dose or acquisition time \cite{spectrum2018}. 

\begin{figure}[ht!]
    \centering
    \includegraphics[width=1\textwidth]{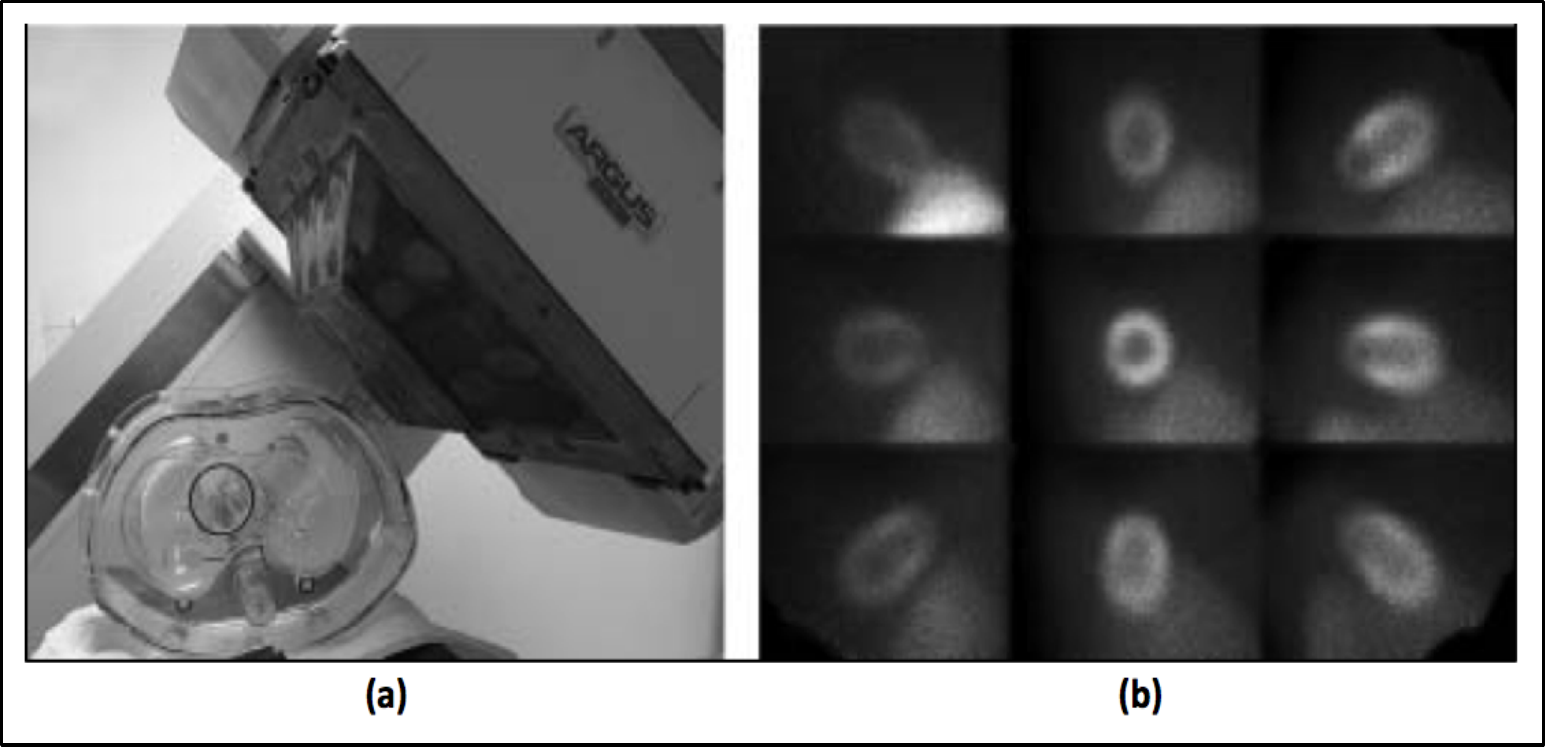}
    \caption[Multi-pinhole gamma camera design.]{Multi-pinhole gamma camera design. (a) 9-collimator multi-pinhole gamma camera by Funk. et. al, experimental setup with anthropomorphic phantom (b) 9-projections acquired of the cardiac torso \cite{funk2006novel}.}
    \label{ch1_funk}
\end{figure}

Funk et. al. \cite{funk2006novel} in 2006 proposed another Generation-II cardiac SPECT camera that uses 9 pinhole collimators. It is a gantry static geometry which is almost as sensitive as DSPECT. The experimental setup and the acquired projections for cardiac torso are shown in Figure \ref{ch1_funk}. Another successful Generation-II camera is GE Discovery Nuclear Medicine 530c which again employs gantry static geometry with pinhole collimators and $8\text{cm} \times 8\text{cm}$ flat detectors. Different pinholes are arranged in an arc around body contour. This uses CZT detectors which have much better energy resolution. The system performance evaluation of GE Discovery shows the 3.5-6 times improvement in sensitivity \cite{esteves2009novel}. Dey \cite{dey2013spect, dey2012improvement} proposed a Generation-III cardiac SPECT system that uses multi-pinholes with curved detectors. It is also a static geometry in which pinholes are arranged in an arc around the body contour. Different shapes of curved detectors have been studied analytically and shown to improve the sensitivity over Generation-II cameras. This thesis focuses on studying Generation-III gamma camera with hemi-ellipsoid CsI detectors using Monte-Carlo acquisition. We will be comparing the performance of our system with GE Discovery in literature \cite{kennedy20143d}.

\section{Problems and Motivation}
\label{sec17}
The necessity for collimation is the key feature that separates SPECT from other nuclear imaging modalities. The characteristics of collimator is one of the deterministic factors of spatial resolution and system sensitivity as described in previous sections in the introduction. Despite many changes made to the SPECT system in last couple of decades, it suffers from low sensitivity. Resolution of the state of art system (e.g. GE Discovery) is not so great \cite{kennedy20143d}. Improving sensitivity without degrading the spatial resolution is the major challenge in SPECT imaging.
	
Cardiac SPECT is a crucially important non-invasive imaging modality. Every year $\sim7$ million patients in US go through cardiac scan using SPECT for diagnostic purposes in order to assess myocardial perfusion and related cardiac health risks, and the number is much larger worldwide. Since, it uses ionizing radiation for imaging, radiation exposure to the patient is of course a concern. A comparative study of radiation exposure due to different diagnostic imaging systems has shown that the nuclear medicine is the second highest contributor of radiation dose to the public, and half of which comes from the cardiac SPECT. CT is on the top of the list for radiation exposure to the general population \cite{hall2006radiobiology, einstein2007radiation, duvall2013reduction}.
	
The common protocol for cardiac scan using Generation-I system is two-day; stress study with 25mCi injection is followed by rest study with 15mCi injection the next day. However, the Generation-II cameras have 5-8 times improved sensitivity reducing full doe acquisition times to 2-4 minutes \cite{slomka2009advances, smith2013recent, funk2006novel}. These systems can also be used for so called “stress-first” protocol which are sometimes referred to as “stress-only” protocol. In “stress first” protocol, rest study can be avoided if the stress study looks normal. This helps in reducing unnecessary radiation exposure to the patient. But, vast majority of hospitals are still using Generation-I system as the Generation-II cameras are not yet prevalent. A key point to note here is, Generation-II cameras can be used to reduce radiation dose to the patient significantly (3mCi), but the image acquisition time is still 10-12 minutes \cite{slomka2012advances, funk2006novel}. Longer acquisition times results in patient discomfort leading to motion artifacts in the image. Slower hospital workflow is another issue caused due to longer acquisition time as it makes the service more expensive.
	
Ideally, one would want to significantly reduce both acquisition time and radiation dose to the patient. Third generation camera proposed by J. Dey \cite{dey2013spect} which uses multi-pinhole collimation with curved CsI detectors exactly does that. The fundamental logic behind this is, resolution gets improved near the center of curved detector due to increased magnification. And, the improved resolution can be traded for higher sensitivity by increasing the pinhole diameter to match the resolution of flat detector system (GE like system) \cite{kalluri2015multi}. Meaning, the Generation-III system will have same resolution as state of art system but improved sensitivity. The spatial resolution of GE like system (multi-pinhole collimator with flat detectors) is given by \cite{peterson2012advances},
\begin{equation}
\begin{array}{l}
\operatorname{Res}=\sqrt{d^{2}\left(1+\frac{b}{a}\right)^{2}+\left(\frac{b}{a}\right)^{2} R_{I}^{2}} \\
\text {\hspace{20mm} with} \quad d=\sqrt{d_{0}\left(d_{0}+2 \mu^{-1} \tan \frac{\alpha}{2}\right)}
\end{array}
\end{equation}
where $d$ is the effective pinhole diameter which includes adjustment for pinhole penetration and acceptance angle of collimator. $d_0$ is the actual physical pinhole diameter. $\mu$ is the linear attenuation coefficient of collimator material. $\alpha$ is the acceptance angle. $R_I$ is the intrinsic detector resolution which includes depth of interaction (DOI) component.

Since the spatial resolution for curved detector varies from one point to another, the average value of resolution is calculated for the performance assessment. An analytical expression for average spatial resolution for curved paraboloid detector with pinhole collimation is \cite{dey2012improvement},
\begin{equation}
\operatorname{Res}_{\mathrm{av}}^{2}=
\frac{\int_{0}^{H}\left[\left(1+\frac{b}{a+H-h}\right)^{2} d^{2}+\left(\frac{b}{a+H-h}\right)^{2} R_{I}^{2}\right] \frac{\left(R^{4}+4 R^{2} H h\right)^{1 / 2}}{2 H} dh }{\int_{0}^{H} \frac{\left(R^{4}+4 R^{2} H h\right)^{1 / 2}}{2 H} d h}
\end{equation}
where $H$ is the height of paraboloid detector, $R$ is the base radius of paraboloid detector, and $h$ denotes the variable height at any arbitrary detector point. The advantages of having curved detectors include improved resolution, improved sensitivity, lower acquisition time, and same packing fraction as its flat detector counterpart. Further technical details about curved geometry and its benefits can be found in \cite{dey2012improvement}.

There have been some theoretical studies on performance evaluation of different curved detector geometry. Dey \cite{dey2012improvement} demonstrated 29\% improvement in resolution with paraboloid detector which can be traded for sensitivity gain of 2.25 times.  Similarly, for trapezoidal detector \cite{agarwal2011low}, the sensitivity gains with respect to state-of-art systems were found to be 2.26 times.

\section{Hypothesis and Specific Tasks}
\label{sec18}
This thesis is focused on studying and assessing the performance of third generation gamma camera SPECT system using 21 pinholes and hemi-ellipsoid detectors. The details of geometry that we use is discussed in detail in Section \ref{sec222}. Some earlier analytical simulations on different curved detector based SPECT system has demonstrated the significant improvement in resolution/sensitivity \cite{dey2012improvement, agarwal2011low}. Hypothesis: the 21 pinhole SPECT system using hemi-ellipsoid detector improves the photon detection sensitivity compared to Generation-II systems like DSPECT and GE discovery without worsening the spatial resolution. Following three specific tasks were carried out for the performance assessment or hypothesis testing.
\begin{enumerate}
\item GATE (Geant4 Application for Tomographic Emission) point source simulation to find the high sensitive operating pinhole diameter.
\item GATE simulation and reconstruction of rod phantom (Jaszczak like phantom) to evaluate the spatial resolution throughout the volume of interest (VOI).
\item GATE simulation and reconstruction of NCAT (NURBS based cardiac torso) heart for the proof of concept, resolution quantification, and sensitivity comparison with ‘state-of-art’ system.
\end{enumerate}

\section{Organization of this Thesis}
\label{sec19}
This thesis consists of three major chapters. Chapter \ref{ch1} is introduction in which different medical imaging modalities are introduced. Since, the focus of this thesis is on newly proposed third generation gamma camera SPECT system, background information about gamma camera, different SPECT imaging modalities, and imaging protocols are reviewed. In addition, comparative account of Gen-I and Gen-II gamma camera is presented in another subsection of introduction. Problems with the ‘state of art’ system and what are we proposing to do about it are highlighted in Motivation sub-section. Furthermore, reconstruction algorithms that will be used in the project are discussed briefly. Chapter \ref{ch2} is the body of the thesis which is based on a research article titled “Performance Analysis of a High-Sensitivity Multi-Pinhole Cardiac SPECT System with Hemi-Ellipsoid Detectors” that was submitted to Medical Physics. As of now, the status of the manuscript is ‘conditionally accepted for publication’. The authors of the paper are Narayan Bhusal, Dr. Joyoni Dey, Jingzhu Xu, Dr. Kesava Kalluri, Dr. Arda Konik, Dr. Joyeeta M. Mukherjee, and P. Hendrik Pretorius \cite{bhusal2019performance}. Lastly in Chapter \ref{ch3}, conclusions of the thesis are presented and possible future directions of this research are discussed briefly.

%% file: chapter2.tex
\label{ch2}
\section{Background and Motivation}
Cardiac SPECT is an important non-invasive modality to assess myocardial perfusion, ischemic defects, abnormal heart wall motion, etc., with $\sim$7 million patients/year undergoing nuclear cardiology scans in the USA. However, of all the diagnostic imaging modalities, nuclear medicine is the second highest contributor of radiation exposure to the general public, behind Computed Tomography (CT) \cite{hall2006radiobiology, einstein2007radiation, duvall2013reduction}. Cardiac SPECT contributes about half of this exposure. Standard Anger-camera-based systems in the clinic utilize a 10-12 min $\sim$25mCi stress-study followed by a second-day $\sim$15mCi rest study, spanning 16-20 minutes, leading to patient motion, patient discomfort, and in-efficient hospital workflow. The patient motion may cause misdiagnosis due to motion-induced artifacts in reconstruction and misalignment of transmission and emission reconstructed images \cite{mukherjee2010quantitative, mukherjee2013evaluation, mukherjee2009estimation, dey2015effect}.

A new generation of dedicated Cardiac SPECT systems with improved sensitivity of 3-8 times \cite{chan2016impact, slomka2009advances, esteves2009novel, nakazato2013myocardial, erlandsson2009performance, gambhir2009novel, volokh2008myocardial, blevis2008czt, volokh2008effect, dic2018web} over standard clinical systems has emerged. The sensitivity improvement depends on several factors, such as patient size and activity uptake, field-of-view, and baseline system geometry to compare with. Most of the second-generation dedicated cardiac designs place detectors close to the body, focusing on a region of interest around the heart. Nakazato et al. \cite{nakazato2013myocardial}, Erlandsson et al. \cite{erlandsson2009performance}, and Gambhir et al. \cite{gambhir2009novel}. analyzed the Dynamic SPECT (D-SPECT) system, which uses parallel-hole collimation. The planar sensitivity improvement of D-SPECT, compared to a general-purpose SPECT camera, was 5.5 times, and for tomographic reconstruction the improvement was 4.6-7.9 times for the heart region \cite{erlandsson2009performance}. The acquisition time for clinical studies was 5.5 times shorter (2 minutes for D-SPECT versus 11 minutes for the general-purpose system) \cite{gambhir2009novel}. Nakazato et al. \cite{nakazato2013myocardial} acquired $\sim$8Million LV-counts in 14 minutes with DSPECT, and about 1.13million LV counts in 2mins. A feature of the GE Discovery camera design is that there are no moving parts, thus allowing dynamic SPECT imaging as well as reducing the servicing costs. Esteves et al., \cite{esteves2009novel} studied the GE Discovery Nuclear Medicine 530c (DNM) on 168 patients. The rest and stress acquisition times were 4 and 2 minutes, respectively, for the GE Discovery system and 14 and 12 minutes, respectively, for a standard dual detector SPECT camera (S-SPECT), implying 3.5-6 times sensitivity gain. 
	
The new generational dedicated cardiac systems enable “stress-first” SPECT protocols with lower doses, and obviates the need for subsequent rest-studies if stress-studies are normal ($\sim$60\% of cases) \cite{einstein2007radiation, duvall2013reduction, duvall2014comparison, duvall2012prognosis, des2009stress}. This has been shown to reduce radiation exposures to patients and associated personnel \cite{duvall2013reduction} but acquisitions take about 10-14 min \cite{einstein2007radiation, duvall2013reduction, nakazato2013myocardial, duvall2014comparison, duvall2012prognosis, des2009stress}.  Additionally, these new Cameras are not yet prevalent, with standard Anger-camera based systems still used for the vast majority of patients. 
	
We proceeded to explore if we can design a higher sensitivity Cardiac SPECT system (Dey \cite{dey2013spect, dey2012improvement}) in order to reduce patient exposures and image acquisition times. The main idea is to use curved detectors to improve resolution. The improved resolution can then be traded with improved sensitivity using a larger pinhole diameter \cite{dey2012improvement}. 
	
Dey \cite{dey2012improvement} previously explored a theoretical hemi-paraboloid system with analytical forward system acquisition simulation of point sources, yielding 2.26 times sensitivity improvement over a base flat-detector system for equivalent average FWHM. We did a preliminary exploration of the hemi-ellipsoid detector shape \cite{kalluri2015multi} and estimated that further performance improvement is possible compared to a hemi-paraboloid shape of the same base diameter and height (because of higher magnification in the center over a larger angular sector).  
	
The goal of this work is to rigorously evaluate the resolution and sensitivity of a system with 21 hemi-ellipsoid detectors in reconstruction space, for GATE (Geant4 Application for Tomographic Emission) acquisitions of point/rod sources and NCAT phantom and compare the performances to existing literature on state-of-the-art systems such as GE discovery and DSPECT.

\section{Methods}
\label{sec22}
The main idea behind using a curved detector instead of a flat one for multi-pinhole (MPH) SPECT is explained in a previous work \cite{dey2012improvement}, briefly summarized here. Assuming the pinholes will be close to the body surface for best sensitivity, we show (Figure 3 in manuscript \cite{dey2012improvement}) that once the object depth from pinhole-aperture and angle of acceptance is fixed by application, curved detectors, as opposed to flat detectors, will allow for more detector area and better packing factor for a compact geometry of detectors. An inverted wine-glass shaped detector collimated by pinhole will improve magnification in the central section and improve resolution compared to a flat-detector. The parameters for collimator height \enquote{$a$} were investigated in that work \cite{dey2012improvement} in depth. For this work we used the parameter determined in that paper \cite{dey2012improvement} allowing for large field of view (200mm at depth of 150mm from pinhole, which is approximately our depth of interest for the heart). In this work we investigate a full system with 21 hemi-ellipsoid curved detectors and analyze the performance compared to state-of-the-art clinical systems. The hemi-ellipsoid detector system is termed ellipsoid system for simplicity here onwards.
	
First, we compared the full-width-half-maximum (FWHM) versus pinhole-diameter for a single hemi-ellipsoid detector with pinhole collimation and a single base-flat-detector with the same pinhole collimation, using point sources simulated with GATE. The pinhole diameter was varied over a range. This gives a system geometry-independent \enquote{raw} comparison points, between the two detectors (ellipsoid versus flat). Also importantly, this gives a higher-sensitivity operating point, or a higher pinhole diameter setting for the ellipsoid detector for equivalent average acquisition resolution to the flat-detector. 
	
For this operating point (based on equivalence in average resolution with base flat detector), we performed full system resolution analysis. Resolution analysis requires full-system acquisition and evaluation in reconstructed space. Therefore, in the next step, we obtained GATE acquisition simulations of 21 projections for our hemi-ellipsoid multi-pinhole system for a series of rod sources in our volume of interest (VOI) (similar to GE discovery system resolution evaluation \cite{kennedy20143d}). We compared the FHWM of our system with the GE system. As done for GE discovery evaluation \cite{kennedy20143d}, the collimator blur is compensated in iterative reconstruction. 
	
Finally, in a third step, we obtained GATE acquisitions for the mathematical anthropomorphic NCAT (NURBS-based Cardiac Torso) \cite{segars1999realistic} phantom with a full (clinical) dose acquisition and estimated the LV counts and compared FWHM of LV-wall in the reconstructed images. We also acquired an ultra-low dose acquisition of $\sim$3mCi (as in other low-dose studies \cite{nakazato2013myocardial, duvall2014comparison, duvall2012prognosis, des2009stress}) for the ellipsoid detector system (with a high-sensitive diameter setting) for comparison. Each step and associated sub-steps is explained in details below.

\subsection{GATE Point Sources simulation comparison between a single Ellipsoid and Flat detector, each collimated by a pinhole}
\label{sec221}
\begin{figure}[ht!]
    \centering
    \includegraphics[width=1\textwidth]{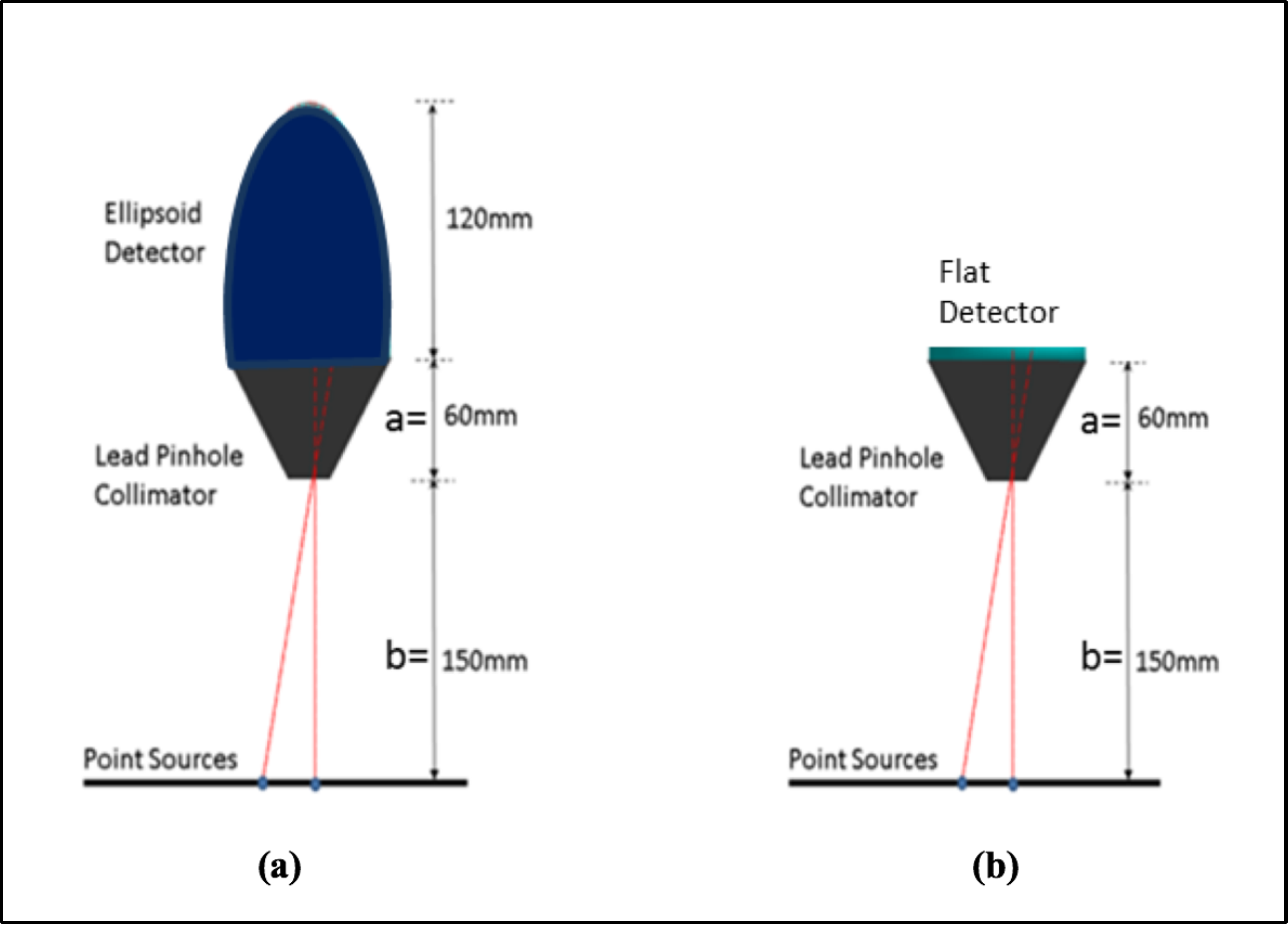}
    \caption[GATE simulation setups.]{GATE simulation setup using (a) Ellipsoid detector and (b) Flat-detector, and point sources located at 150mm from the pinhole aperture.}
    \label{ch2_dets}
\end{figure}

Our scintillator detector design is that of a hemi-ellipsoidal shape (referred to as an Ellipsoid detector) with a CsI crystal of 6mm thickness, 80mm diameter, and 120mm height (Figure \ref{ch2_dets}(a)). For an initial rudimentary resolution-sensitivity analysis (FWHM versus pinhole-diameter), point source simulations were compared between the collimated Ellipsoid crystal in Figure \ref{ch2_dets}(a) and its base flat-detector system in Figure \ref{ch2_dets}(b).

The GATE simulations included Photoelectric and Compton interactions. Only photons detected with energies within a 10\% window around the photo-peak of 140.5 keV are stored. The GATE simulations include pinhole-penetration effects, scatter, and attenuation. All the GATE simulations mentioned in this work were done on a high-performance cluster (HPC) at Louisiana State University.
	
The simulations were obtained for 7 different diameters from 4mm to 10mm, in steps of 1mm, for both the ellipsoid and flat detector. For each pinhole diameter, 9 point sources were placed on a plane 150mm depth below the pinhole diameter at 10mm intervals from the center to the edge at radial distance 80mm. The acquired counts obtained at the detector were binned to 1 $\text{mm}^3$ detector-voxel resolution. The detector-counts were back-projected to a plane at 150mm depth (where the center of the region of interest, the heart, is expected to be located) and FWHM was calculated. 
  	
We plotted the average FWHM (average of the FWHM of the 9 point sources evaluated from the center to the edge of the detector) versus pinhole-diameter, as well as the sensitivity versus average FWHM. These plots allowed us to extract the higher pinhole-diameter setting obtainable for the ellipsoid system for similar average acquisition resolution as a flat-detector with 5mm pinhole diameter. This analysis provides us a higher-sensitive pinhole-diameter operating point for our system.
	
The full system resolution is to be determined in reconstruction space after collimator resolution recovery. In the subsequent sections we describe our system configuration geometry with 21 of these detector-pinhole units spatially arranged around the region of interest, and our GATE evaluation of the full system using arrays of rod-sources in region of interest and compared to the GE discovery system \cite{kennedy20143d}.

\subsection{Configuration Geometry and Reconstruction Algorithm}
\label{sec222}
\begin{figure}[ht!]
    \centering
    \includegraphics[width=1\textwidth]{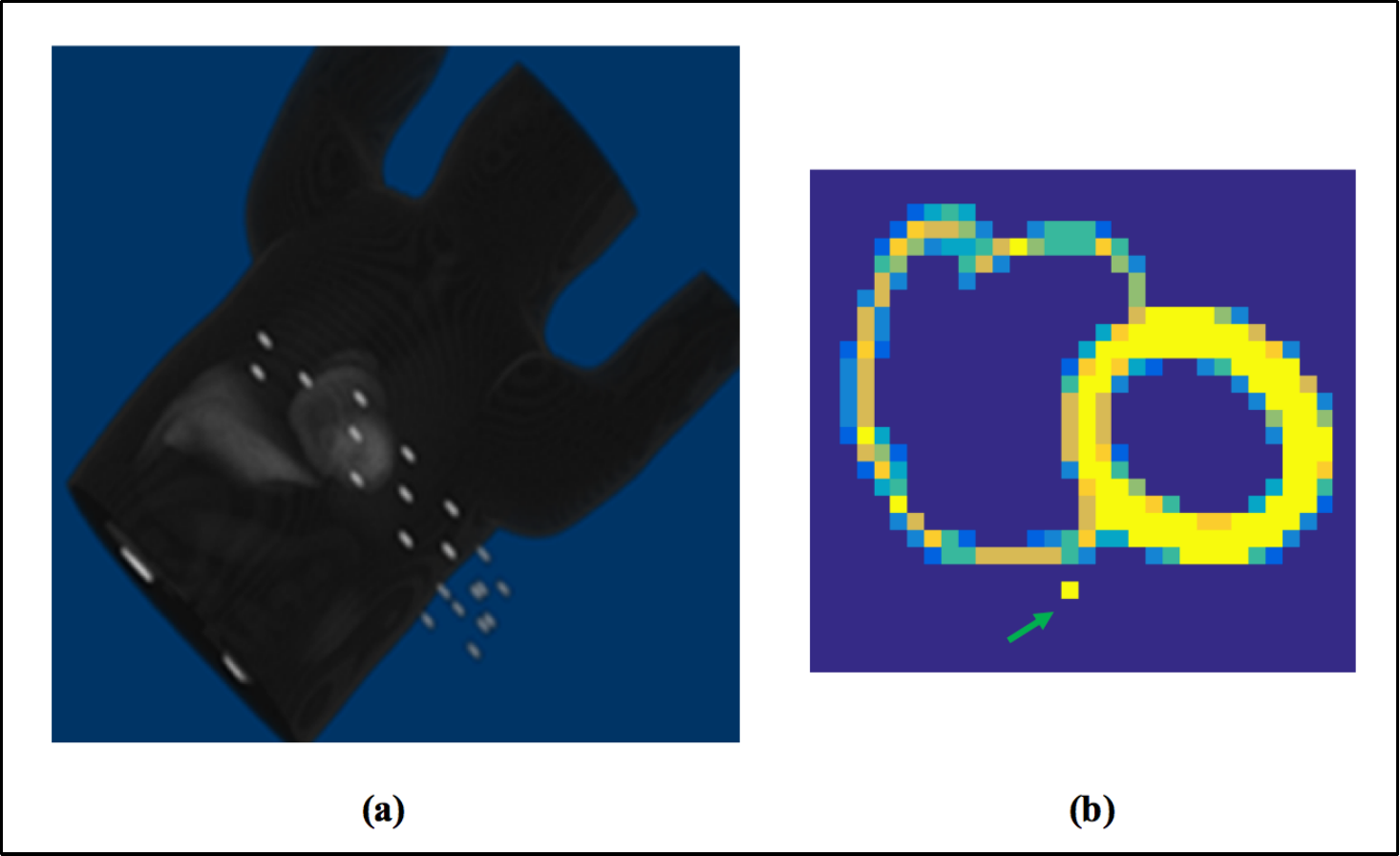}
    \caption[Geometry of our gamma camera SPECT system.]{Geometry of our SPECT system. (a) 21-detector Cardiac-SPECT systems with the NCAT phantom with pinhole diameters arranged in three arcs on a sphere with center beyond the heart, shown in (b). The pinhole-axes meet at the point (called the “iso-center”) indicated by arrow, 200mm below their diameters. Note in (b) the liver is omitted for better visualization.}
    \label{ch2_geom}
\end{figure}

Geometry:  A stationary 21-pinhole configuration geometry with pinholes respectively distributed on 3 arcs of a spherical surface is shown in Figure \ref{ch2_geom}(a). The top arc has 6, the middle arc (most sensitive zone) has 9, and the last arc has 6 pinhole-detector units. The geometry was determined heuristically: it was ensured that the NCAT heart region is well within the FOV and each detector-pinhole unit is able to image the entire heart without truncation. All pinholes’ central axes point towards the heart region such that they converge to a point at a distance of 200mm below the surface, beyond the heart on the NCAT phantom shown in Figure \ref{ch2_geom}(b). This is called the “iso-center” of the geometry for convenience.

We tested this geometry in GATE simulations of rod sources and NCAT, with the ellipsoid-detectors (called Ellipsoid-detector system) mounted on pinholes.  We considered two settings of pinhole: (1) a 5mm diameter pinhole similar to GE Discovery \cite{kennedy20143d}, expected to achieve a clinical level of counts. (2) the high-sensitivity setting of pinhole diameter, determined by analysis of imaging point sources with singleton detector-pinhole units described in Methods Section \ref{sec221}. While we will show the analysis later (in Results), for clarity of presentation, we mention the high-sensitive diameter was determined to be 8.68mm. 
	
The GATE system simulations of NCAT and rod-sources took over 500K CPU hours (and over 6 months) in the HPC cluster.
	 
MPH Reconstruction: A multi-pinhole MLEM/OSEM reconstruction algorithm developed by Dey \cite{chan2016impact, kalluri2015multi, dey2011comparing} was used to reconstruct the rod-sources and the NCAT phantom acquired in simulations by GATE. The sampling was voxel-based (ray-casting based regular sampling of each voxel). The algorithm compensated for collimator resolution, pinhole sensitivity and attenuation due to intervening body-tissue \cite{chan2016impact, kalluri2015multi, dey2011comparing}. The collimator resolution was compensated by sampling of the pinhole \cite{feng2010modeling}. The pinhole-diameter sampling interval was 0.38mm in two directions. The NCAT phantoms were reconstructed using OSEM by choosing subsets of 3 from 21 projections. The approximate speed-up between MLEM and OSEM was a factor of 6.

\subsection{Resolution Comparison to GE Discovery System: Multiple (21) detector-pinhole GATE Rod-Source Ellipsoid-detector System Acquisitions and Reconstructions}
\label{sec223}
Following the methodology for evaluation of the GE discovery system \cite{kennedy20143d} for a fair comparison, we imaged a rod-source phantom with background activity and reconstructed the images with collimator resolution recovery. We evaluated the FWHM in 3D at the reconstructed rod-sources and interpolated over 3D volume to obtain the FHWM over the entire volume of interest (VOI). The VOI was an oval of dimension 200mm, 180mm, and 180mm such that NCAT heart voxels were well inside the VOI. The rod-sources were of diameter 1mm and length 2mm, spaced 30mm in each direction. Radioactivity of 2MBq was simulated for each rod-source. The acquisition was performed for Ellipsoid-system with 5mm pinhole diameter (clinical sensitivity), as well as the 8.68mm diameter pinhole (high sensitive setting, determined by Methods Section \ref{sec221}). The detector binning was 3mm in each direction.
	
The images were then reconstructed using MPH MLEM reconstruction. The reconstruction voxel size is 2mm in each direction. FWHM was estimated in X, Y, Z, and the worst case of these ($\text{FHWM}_\text{WC}$) was noted. The values were tri-linearly interpolated to obtain the $\text{FHWM}_\text{X}$, $\text{FHWM}_\text{Y}$, $\text{FHWM}_\text{Z}$, $\text{FHWM}_\text{WC}$ at every point on the VOI. To compare with the GE system presentation, the interpolated values of these 4 parameters were shown in the central axial and central coronal slices for an iteration where values have more or less converged. Additionally, we presented the information in the sagittal slice. We also presented the overall-average (over all acquired rod-source points) and the standard-deviation across iterations.

\subsection{GATE NCAT Simulations comparison between Ellipsoid and Flat detector systems}
\label{sec224}
To simulate a realistic uptake of Tc-99m in the heart, liver, lungs and background in GATE, source phantoms for each organ were created separately using the NCAT software. The heart, liver, lung and background relative activities were 100:50:5:10. An attenuation map for NCAT was also generated. For a full injected dose of 25mCi, the uptake in the heart source phantom is assumed to be 0.5mCi (which is, about a 0.3mCi, or 1.2\% in the LV region) \cite{dey2009theoretical}. Therefore, the activity per-voxel is scaled such that a total of 0.5mCi was simulated in the heart-region voxels. Each of the 21 projections was obtained by acquiring the data for 120secs. The three organs (heart, liver, lungs) and the background were acquired in parallel. A 72-hour wall time on the HPC cluster required the division of each simulation into smaller units of time and activity. For example, for the liver, three sets of activity and 12 sets of time (10 secs each) were required. 
	
The GATE events detected by each of the 21 CsI detectors from the different organs were added and binned into detector voxels of 3mm size. These projections were the “measurement” inputs for the MLEM reconstruction algorithm to obtain the final reconstructed image. The full dose data was acquired for Ellipsoid detector systems with 5mm (referred to as Ell5mmFD) and 8.68mm diameter pinholes (Ell8.68mmFD). The Ellipsoid 8.68mm pinhole diameter was also obtained for low dose of 3mCi (consistent with clinical protocols \cite{einstein2007radiation, duvall2013reduction, duvall2014comparison, duvall2012prognosis, des2009stress}). Since the sensitivity is about 3.06 times higher and the dose was reduced 8.3 times, the acquisition time was increased to 5.44 mins (2min x 8.33/3.06) to get similar level of counts. The ultra-low-dose acquisition is referred in short as Ell8.68mmULD. Note that since the acquisition counts approximately linearly scales with the input Bq per voxel and with time, this acquisition data can be equivalently thought of as an ultra-fast (39.2 sec) acquisition at full dose (25mCi injected, 0.5mCi in heart area). Alternately, this case can be thought of as a 2min acquisition with $\sim$8.2mCi injected dose. 
	
Overall Left Ventricle Sensitivity: The heart-only counts for the system were acquired for all three acquisitions and corrected for LV-only and compared to DSPECT data available in the literature \cite{nakazato2013myocardial}. 
 	
 Resolution Analysis on NCAT Reconstructions: The all-organ-acquisitions were reconstructed using the MPH-OSEM, with 4.67mm resolution voxel size. FWHM analysis was done on the short-axis slices before application of clinical smoothing, similar to methods in literature \cite{el2000relative}. The LV intensity was extracted in different profiles around the short-axis slices. To reduce effects of noise, each profile consisted of the average of three neighboring profiles. Four profiles, two vertical (superior and inferior) and 2 horizontal (anterior and posterior), were extracted for 10 short axis slices. The corresponding profile from the corresponding short-axis slice of the oriented NCAT phantom was extracted. The normalized NCAT profile was convolved with a Gaussian and the best fit of the resulting signal to the normalized reconstructed profile was found iteratively using Matlab (Mathworks, MA) function fmincon. The normalization was important to eliminate the effect of any reconstruction bias. The FWHM of the best fit Gaussian was found and the average FWHM (of four profiles) for each slice was calculated. Rather than tabulating FWHM for all 10 slices, we further averaged over three slices for each of the following three regions: mid-short-axial region, towards base and towards the apex and tabulated the average regional results for the three different systems: Ell5mmFD, Ell8.68mmFD and Ell8.68mmULD. 
	
Short, Long Axes and Polar map: The reconstructed datasets from the GATE acquisitions were clinically smoothed \cite{chan2016impact, kennedy20143d} and displayed in short and two long axis slices as well as polar maps for ELL5mmFD, ELL8.68mmULD and ELL8.68mmFD. The original NCAT was also similarly smoothed and polar mapped for comparison.  
	
Bias and Variance: Using GATE for large-scale simulations for noise-analysis is prohibitive. Hence, we performed bias-variance analysis with analytical forward simulations and reconstruction. While the analytical method does not estimate the scatter, for Tc99m the scatter is expected to be relatively low \cite{narayanan2003human}. Poisson noise was added (similar to past work \cite{dey2009theoretical, dey2010estimation}) to near noise-less analytical projections. Twenty noise-realizations were reconstructed with 4.67mm voxel size. Bias-variance for the three systems was plotted.

\section{Results}
\label{sec23}
\subsection{GATE point source simulations: Comparison for a single pinhole-collimated Ellipsoid and Flat detector}
\begin{figure}[ht!]
    \centering
    \includegraphics[width=1\textwidth]{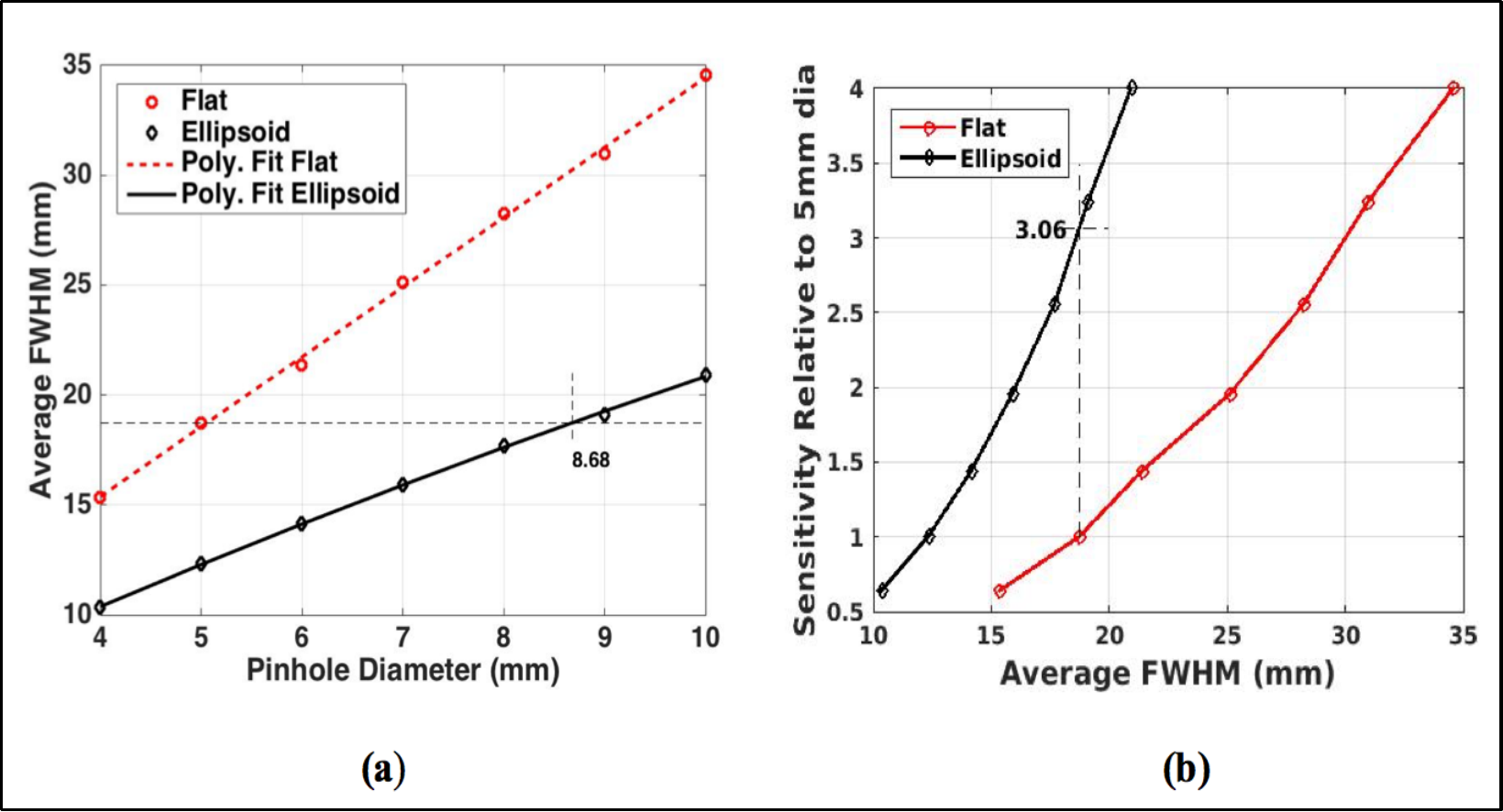}
    \caption[Plots of GATE simulation results, showing raw resolution comparison.]{Plots of GATE simulation results: (a) Average FWHM plotted against pinhole diameter. Ellipsoid case is interpolated to show that, for same average resolution for 5mm diameter for the Flat detector case, an 8.68mm diameter may be used for the Ellipsoid case. (b) Sensitivity with respect to 5mm diameter case (i.e., $d^2/25$) is plotted versus the average FWHM. These imply a 3.06 times sensitivity improvement.}
    \label{ch2_rawres}
\end{figure}

Figure \ref{ch2_rawres}(a) shows the average FWHM versus pinhole-diameter for the 7 diameter settings with flat and ellipsoid detectors. We immediately see that the FWHM increases at a steeper rate (therefore faster loss of resolution) with pinhole-diameter for the Flat detector compared to the Ellipsoid detector. The polyfit interpolation (MATLAB, Mathworks, MA) to fit the data is also displayed, showing a linear-trend for flat and ellipsoid. Figure \ref{ch2_rawres}(b) plots the data as sensitivity (normalized versus the 5mm-pinhole-diameter) versus average FWHM. The relationship is nonlinear (approximately quadratic) for both, with the sensitivity showing steeper rate of improvement for the Ellipsoid detector. As detailed later, the 5mm setting Ellipsoid acquired a clinical level of counts in GATE for the NCAT phantom. We see in Figure \ref{ch2_rawres}(a) that at 8.68mm diameter, the Ellipsoid detector system had similar raw acquisition resolution as the 5mm Flat detector system, at the depth of 150mm (center of region of interest). Similarly, from Figure \ref{ch2_rawres}(b), for the same average resolution for the Ellipsoid detector, we expect about a 3.06 times sensitivity improvement with respect to the Flat detector with 5mm pinhole diameter. 

This provides us with an operating point of 8.68mm diameter for the Ellipsoid detector system for further studies with a point source and NCAT phantom and allows us to investigate system resolution after reconstruction with the collimator resolution recovery. In the next section, we will compare the Ellipsoid 8.68mm with GE Discovery FWHM reported in the literature.

\subsection{Rod Source Resolution Analysis Post-Reconstruction and Comparison to GE Discovery System}
\begin{figure}[!ht]
    \centering
    \includegraphics[width=0.8\textwidth]{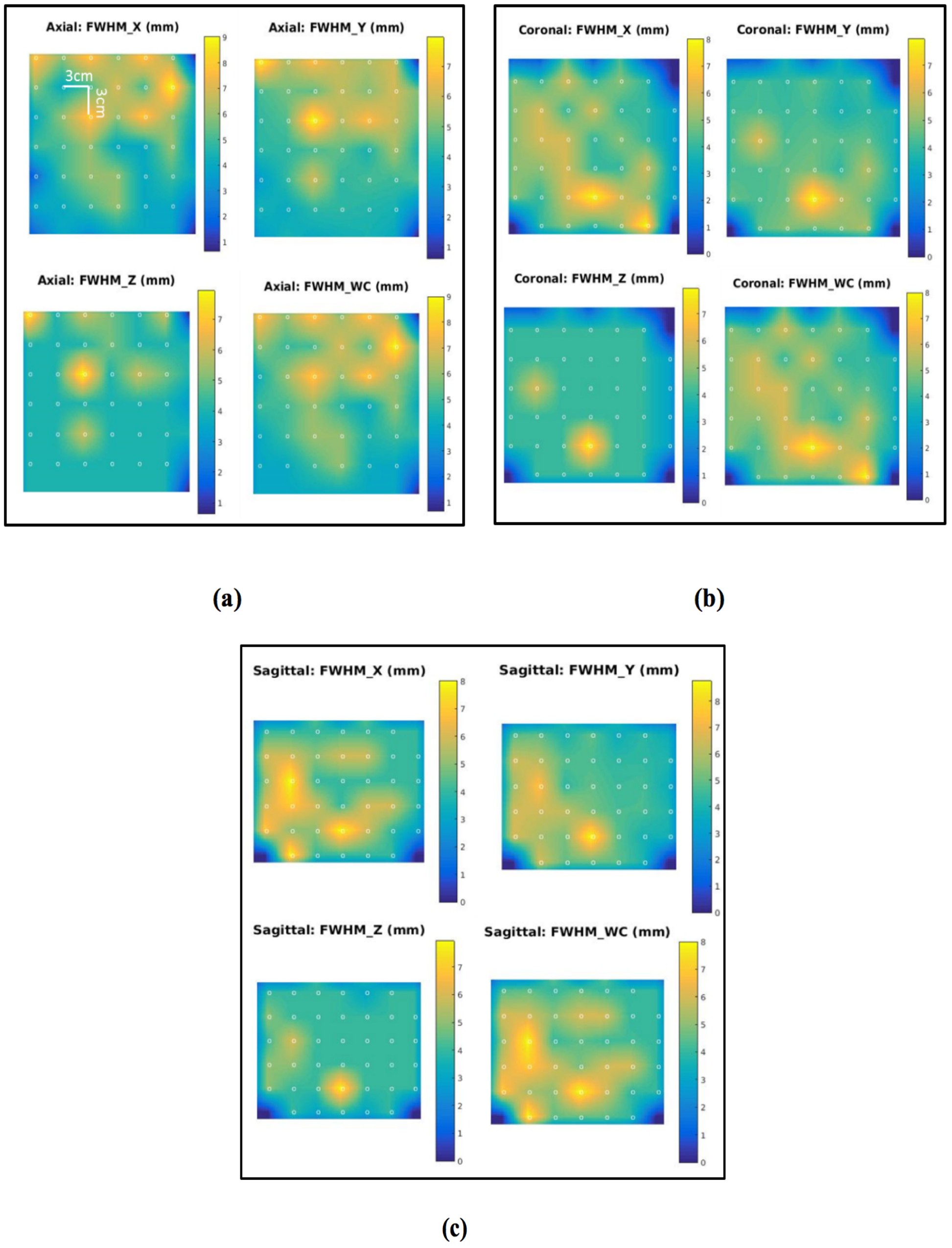}
    \caption[FWHM (mm) for x-y-z and the worst-case for ELL8.68mm.]{FWHM (mm) is shown for x-y-z and the worst-case for ELL8.68mm system for the 70th iteration of reconstruction. Images show interpolated values for (a) mid-axial slice (b) mid-coronal and (c) mid sagittal slices. The dots represent the acquisition points (spaced 30mm apart in each direction).}
    \label{ch2_fwhms}
\end{figure}

The array of rod-sources in the volume-of-interest (VOI) was imaged, reconstructed with collimator resolution recovery and FWHM extracted as described in Methods Section \ref{sec223}. To compare with the GE Discovery system \cite{kennedy20143d}, we show the interpolated FWHM values (X, Y, Z and worst case WC) in Axial, Coronal and Sagittal slices, in Figure \ref{ch2_fwhms} (a-c) respectively.
\begin{figure}[!ht]
    \centering
    \includegraphics[width=1\textwidth]{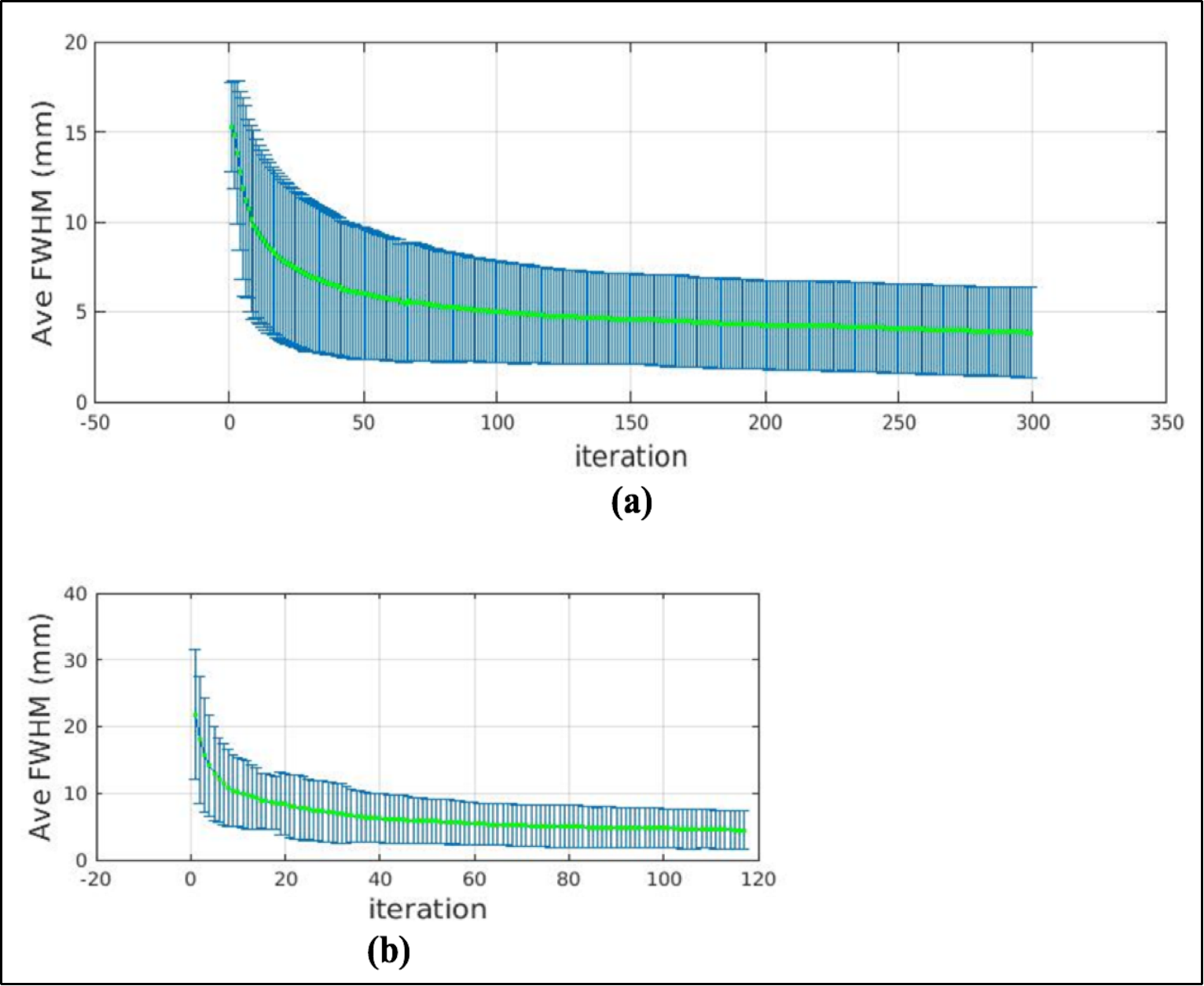}
    \caption[Average overall resolution plotted against the iteration number.]{Average overall FWHM (over VOI) with standard deviation error bar plotted with respect to iteration (a) Ell5mm (b) Ell8.68mm.}
    \label{ch2_resol}
\end{figure}

The average FWHM versus iteration in Figure \ref{ch2_resol} shows ELL5mm converges to similar values as ELL8.68mm but slower. At 300 iterations the convergence is less than 1.5\% (measured by percent difference at each iteration from mean of last 10 iterations), while similar results are achieved for ELL8.68mm at around 118 iterations. The slower convergence for the higher-acquired resolution case (ELL5mm) is expected since resolution recovery typically takes longer for a source acquired with higher resolution setting compared to a lower resolution acquisition. Also, expectedly, the final values after resolution recovery are similar for ELL5mm and ELL8.68mm, with the ELL8.68mm case converging at slightly higher values of FWHM. The one-standard-deviation error bars are shown on the respective plots.

The average values and standard-deviation in X, Y, Z over all the acquired points in the VOI are shown in Table 2.1 for the ELL5mm and ELL8.68mm (at 300th and 118th iterations respectively). Compared to GE Discovery results \cite{kennedy20143d} the FWHM are, in general significantly lower for the Ellipsoid detector system. The overall average for the ELL8.68mm system is 4.44mm as opposed to 6.9mm reported for the GE Discovery system \cite{kennedy20143d} indicating the higher resolution in addition to higher sensitivity of our proposed system.

\begin{table}[!ht]
\label{table2p1}
\caption{FWHM for reconstructed rod sources}
\vspace{2mm}
\centering
\begin{tabular}{|c|c|c|c|c|}
\hline System & \multicolumn{3}{c|} {Average FWHM} &\\
\cline { 2 - 5 } & X & Y & Z & Overall\\
& mm & mm & mm & mm \\
\hline Ell5mm & $4.21(1.42)^{*}$ & 3.82(1.49) & 3.49(1.41) & 3.84(2.49) \\
\hline Ell8.68mm & 4.84(1.68) & 3.97(1.95) & 4.52(1.21) & 4.44(2.84) \\
\hline
\end{tabular}
\\
*quantities in brackets are the standard deviations
\end{table}

Post collimator-resolution recovery, the FHWM of ELL5mm was similar to ELL8.68mm with the former having slightly lower overall FWHM at 3.84mm.  Note the FHWM analysis of the rod-sources is limited by the 2mm voxel size of the reconstructed datasets.

\subsection{NCAT Acquisition and Reconstruction}
Full system 21 projections for Ellipsoid 5mm and Ellipsoid 8.68mm (ELL5mmFD and ELL8.68mmFD) were acquired for 2mins assuming a full injected dose of 25mCi (or 0.5mCi in the heart region). The Ellipsoid system with 8.68mm pinhole diameter was also acquired for 5.44mins assuming 3mCi injected dose, or 0.06mCi in the heart region (ELL8.68mmULD). Photon count details are shown in Table 2.2.

\begin{table}
\label{tab2p2}
\caption{Acquired system counts for NCAT in GATE. The abbreviations Ell5mmFD = (Ellipsoid 5mm, full-Dose, 2mins), Ell8.68mmULD = (Ellipsoid8.68mm with 8.33 times less dose, 5.44mins), Ell8.68mmFD = (Ellipsoid 8.68mm, Full dose, 2min).}
\centering
\resizebox{\columnwidth}{!}{\begin{tabular}{|c|c|c|c|c|c|c|}
\hline System & \multicolumn{2}{c|} {All Organs} & \multicolumn{2}{c|} {Heart-only-Counts} & \multicolumn{2}{c|} { Estimated LV-counts }\\
\cline {2 - 7} & Total Counts & Ave Counts/proj & Total Counts & Ave Counts/proj & Total Counts & Ave Counts/proi\\
\hline Ell5mmFD & $5.99 \mathrm{M}$ & $285.40 \mathrm{K}$ & $2.30 \mathrm{M}$ & $109.67 \mathrm{K}$ & $1.37 \mathrm{M}$ & $65.18 \mathrm{K}$ \\
\hline Ell8.68mmULD & $5.42 \mathrm{M}$ & $258.20 \mathrm{K}$ & $2.08 \mathrm{M}$ & $99.28 \mathrm{K}$ & $1.24 M$ & $59 \mathrm{K}$ \\
\hline Ell8.68mmFD & $13.14 \mathrm{M}$ & $625.51\mathrm{K}$ & $6.38 \mathrm{M}$ & $303.81\mathrm{K}$ & $3.79 M$ & $180.67\mathrm{K}$\\
\hline
\end{tabular}}
\end{table}

Table 2.2 shows the all organ counts (from liver, heart, lungs), just the heart-counts, and the estimated LV counts for the three systems. The LV counts are estimated to be 59\% of the heart (based on the ratio of the sum of the activity for the LV and that of Heart voxels of the NCAT phantom). Extrapolating from data for a full-dose 14 min acquisition \cite{nakazato2013myocardial}, a 2min acquisition for DSPECT will produce $\sim$1.13MC (million counts) in the LV. Thus, our results indicate that the ELL5mmFD (ellipsoid system with pinhole diameter 5mm and full injected dose) have sensitivities slightly better than or comparable to the DSPECT \cite{nakazato2013myocardial}. For ELL8.68ULD (ellipsoid system with pinhole diameter 8.68mm and ultra-low injected dose of 3mCi), the LV counts are $\sim$1.23M, is slightly higher, than the DSPECT \cite{nakazato2013myocardial}, one of the most sensitive systems currently. For ELL8.68mmFD (full dose 8.68mm pinhole aperture) the LV count was 3.79M or about 3.35 times higher than DSPECT case.

\begin{table}[!ht]
\label{table2p3}
\caption{NCAT short-axis slice FWHM analysis (after 12 OSEM iterations).}
\centering
\begin{tabular}{|l|c|c|c|}
\hline \text{ } & \multicolumn{3}{|c|} { Ave FWHM (mm) }\\
\cline { 2 - 4 } & Ell5mmFD & Ell8.68mmULD & Ell8.68mmFD\\
\hline Mid-Short-Axial & 3.83 & 5.66 & 4.95\\
\hline Near-Base & 4.30 & 5.74 & 5.62\\
\hline Near-Apex & 4.37 & 6.39 & 6.69\\
\hline
\end{tabular}
\end{table}
	
Table 2.3 shows the FWHM analysis (explained in Methods Section \ref{sec224}) on NCAT short-axis slices for the three systems. Mid-short-axis slice was the average FWHM over 3 slices around and including the mid-axial slice and 4 profiles each. Note that before the Gaussian fit, each profile sums 3 adjacent lines to reduce noise. Similarly, the values are obtained for the base region and the apex region. Note the FHWM analysis of the NCAT reconstructions is limited by the 4.67mm voxel size of the reconstructed datasets. We observe that these FWHM values are consistent with those obtained with rod-sources, if slightly higher as expected with the higher voxel size of reconstruction, etc.

\begin{figure}[!ht]
    \centering
    \includegraphics[width=1\textwidth]{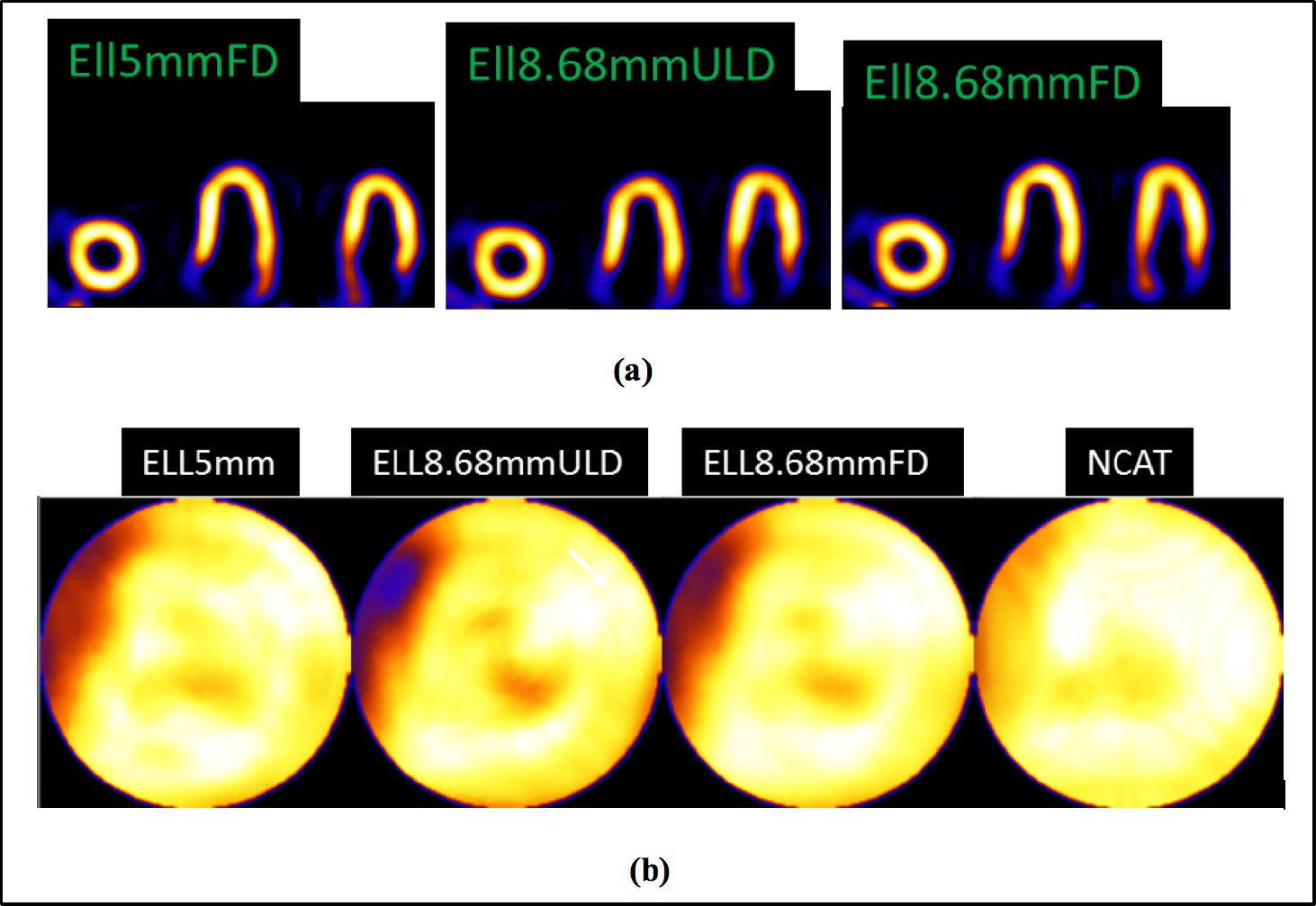}
    \caption[Short-axis and long-axis slices and corresponding polar maps.]{Short-axis and long-axis slices and corresponding polar maps. (a) Reconstructed and re-oriented slices after 12 OSEM iterations and clinical levels of smoothing for Ell5mmFD (full dose, 5mm diameter), Ell8.68mmULD (ultra-low-dose, 8.68mm diameter) and Ell8.68mmFD (full dose, 8.68mm diameter). (b) Polar maps are shown for smoothed Ellipsoid detector systems and NCAT phantom smoothed by the same amount. All (including smoothed NCAT) have the septal wall thinning (white arrow) as present for other geometries and reconstructions \cite{dey2015effect, feng2010modeling}. Note that the mapping procedure expands the base region, spreading out small artifacts.}
    \label{ch2_polmap}
\end{figure}

\begin{figure}[!ht]
    \centering
    \includegraphics[width=0.75\textwidth]{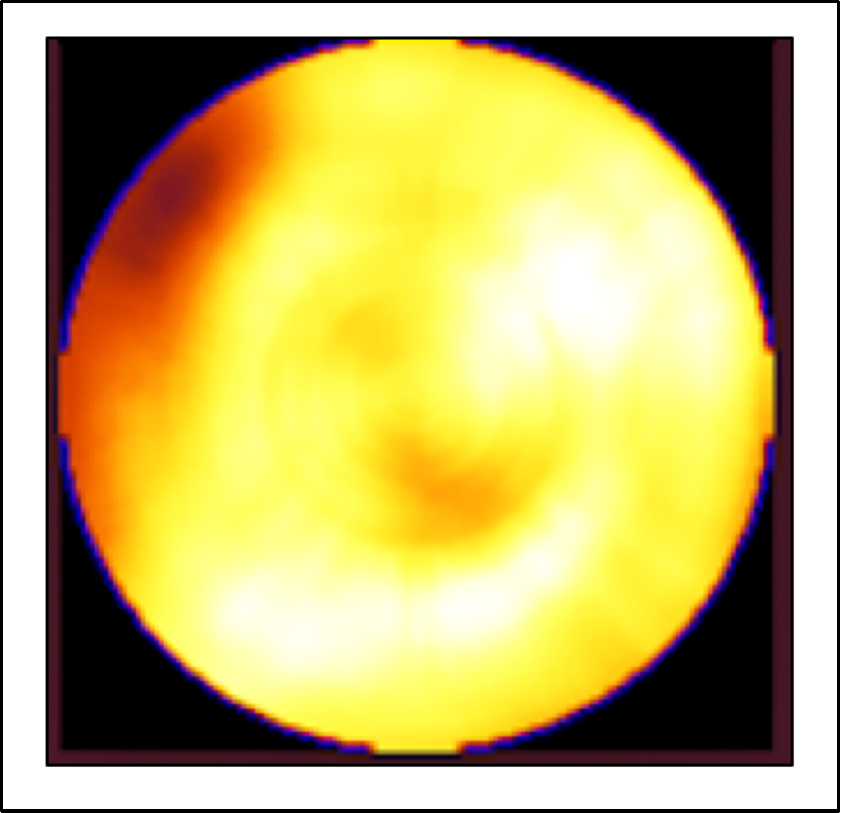}
    \caption[Polar map for ELL8.68mmULD case with 2x2 voxel subdivision in the reconstruction. Clinical level of smoothing applied.]{Polar map for ELL8.68mmULD case with $2\times2$ voxel subdivision in the reconstruction. Clinical level of smoothing applied.}
    \label{ch2_polmap2}
\end{figure}

The short axis and long axes slices are shown at 12th   OSEM iteration in Figure \ref{ch2_polmap} (a) after applying a clinical level of smoothing \cite{chan2016impact, kennedy20143d}.  The polar maps are also shown in Figure \ref{ch2_polmap} (b). The NCAT phantom is similarly smoothed and its polar map is shown for comparison.  Note the septal and apical cooling (present for all the cases including to a small extent, the smoothed NCAT) are due the well-known wall-thinning of the NCAT phantom, present for other system reconstructions \cite{dey2009theoretical, narayanan2003human}. The nature of the polar mapping operation expands the base-region septal artifact in polar-maps. The Ellipsoid systems follow the shape of the NCAT phantom well and ELL8.68mmFD shows least noise and best match overall to smoothed NCAT. However, the basal cooling artifact in the polar map (Figure \ref{ch2_polmap} (b)) can be minimized by using the higher sub-voxel subdivision in the reconstruction algorithm. Figure \ref{ch2_polmap2} shows the polar map for ELL8.68mmULD case with $2 \times 2$ voxel subdivision. The downside to doing voxel subdivision is that it is computational expensive. The computational time increases multiplicatively.

\begin{figure}[!ht]
    \centering
    \includegraphics[width=1.0\textwidth]{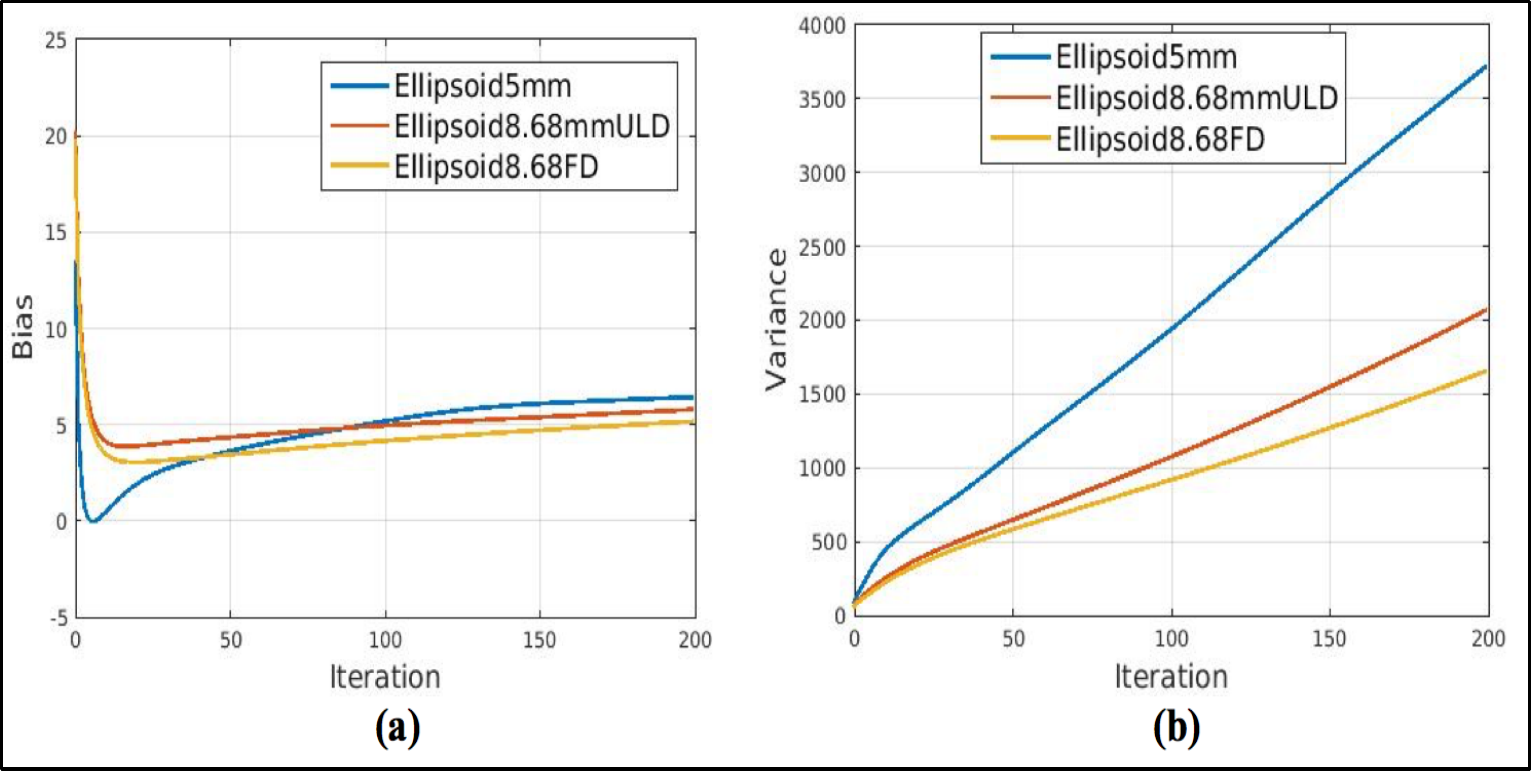}
    \caption[Bias and variance plotted against iteration numbers.]{(a) Bias versus iterations (b) Variance versus iterations of the three systems ELL5mmFD, ELL8.68mmULD and ELL8.68mmFD.}
    \label{ch2_bias}
\end{figure}

Gate acquisitions included pinhole-penetration effects. However analytical simulations showed that given our collimator geometry (annular lead cone of $\sim$1cm thickness) penetration through the pinhole was negligible ($<1\%$) and first order correction showed imperceptible changes in the quality of reconstructed images.  
Finally, bias-versus iterations and variance-versus-iterations are shown in Figure \ref{ch2_bias} for analytical forward simulations and the MPH iterative reconstruction with resolution recovery. The biases roughly converge as expected due to resolution recovery. ELL5mmFD was noisier than ELL8.68mmULD, even though they have similar counts. This can be potentially explained by the lower-resolution acquisition for ELL8.68mmULD and resolution-recovery.  This is consistent with the slightly higher resolution for ELL5mmFD for rod sources and short-axis slices for NCAT.

In summary, the Ellipsoid-detector system setting with 8.68mm diameter pinholes, can acquire similar counts as a clinical system for an ultra-low-dose injection of 3mCi in 5.44mins or 3.35 times higher counts for full-dose 2mins. The rod-source analysis shows an average of $\sim$4.44mm resolution within VOI for the ellipsoid detector system with 8.68mm diameter pinholes. 

\section{Discussion and Future Work}
We showed (Table 2.1-2.3) that the Ellipsoid-detector system setting with 8.68mm diameter pinholes, achieves a higher-sensitivity as well as better resolution than state-of-the-art systems. Note that the system has stationary arrangement of detector-pinhole units, thus can be used for dynamic SPECT imaging where the additional sensitivity will be useful.

Since our system configuration geometry (arrangement of 21 detector-pinholes) and other system parameters are different from clinical MPH GE Discovery system, we compared the FWHMs of the ellipsoid system (with different pinholes and dose/time) directly with a clinical GE Discovery system in reconstruction space with the comprehensive data available from the literature.  

For the purpose of this work, the depth-of-interaction is assumed to be resolved (to within 3mm or half the crystal thickness \cite{bettiol2016depth}), and GATE events are binned to 3mm voxel-size detector. The curved nature of the detector promises to be helpful in lateral as well as depth positioning \cite{dey2017point}. We are building a GEANT4-based look-up-table (LUT) algorithm to recover the depth of interaction for a possible light-readout for a system.

\section{Conclusion}
\label{sec25}
Our Monte-Carlo simulation studies and reconstruction suggest that using (inverted wine-glass sized) hemi-Ellipsoid detectors with pinhole collimators can increase the sensitivity about 3 times over the new generation of dedicated Cardiac SPECT systems (and more than an order of magnitude over standard clinical systems) with average system resolution at 4.44m m over the volume of interest, after resolution recovery in reconstruction.  The extra sensitivity may be used for ultra-low-dose imaging (3mCi) at $\sim$5.44 min, or have an ultra-fast full-dose acquisition in less than 40secs, potentially benefiting millions of patients. Also the stationary geometry and fast acquisition will allow for dynamic imaging where the extra sensitivity will be particularly useful.

%% file: chapter3.tex
\label{ch3}
A novel gamma camera design with 21 pinholes and hemi-ellipsoid CsI detectors is investigated. The data(projections) were acquired using a GATE (Geant4 Application for Tomographic Emission) Monte-Carlo simulation package. A comparative study of FWHM on simple backprojected images for flat detector system (GE like system) and hemi-ellipsoid detector system demonstrated that the 8.68mm pinhole diameter with hemi-ellipsoid detectors will have a similar spatial resolution as its flat detector system with 5mm pinhole opening. However, the gamma photon detection sensitivity of the ellipsoid 8.68mm system is $\sim$3.06 times that of the flat detector counterpart. 

For the performance evaluation of our system, we simulated rod phantom (like Jaszczak Phantom) and investigated the spatial resolution of our system with an 8.68mm pinhole diameter and hemi-ellipsoid detector. The average resolution was compared with the resolution of the GE discovery system. After the image reconstruction, the average spatial resolution over the entire VOI (volume of interest) of our system is 4.44mm. Whereas, the resolution of the GE Discovery system, reported by J. A. Kennedy et. al, is 6.9mm \cite{kennedy20143d}. On the other hand, the sensitivity improvement achieved by this novel design over the state-of-art system is approximately 3 times. 

We also investigated NCAT phantom for proof of principle. Images reconstructed using the OSEM algorithm and corresponding polar maps (displayed in Figure 2.6 of chapter 2) demonstrate the robustness of our system. In addition, we also demonstrate that this system can be used for ultra-low-dose imaging with the injected activity of 3mCi. The spatial resolution of ultra-low-dose imaging (8.33 times less dose and 5.44-minute acquisition) at the mid axial slice of the heart is 5.66mm. 

To summarize, the proposed Generation-III camera with 21 pinholes and ellipsoid detectors has been demonstrated to have $\sim$3 times better sensitivity than Generation-II dedicated cardiac SPECT systems like the DSPECT and GE discovery system. The sensitivity gain, if compared with the Generation-I system, is an order magnitude or more. In addition to better sensitivity, our system has been shown to have better resolution than state-of-art systems. The improved sensitivity can be very beneficial to reduce the radiation dose to the public as it allows for an ultra-low-dose cardiac scan. The sensitivity gain can also be traded for the reduction in acquisition time as it allows for 39.2-sec data acquisition with a full 25mCi injection. The ultra-fast acquisition (39.2 sec) could reduce the problems of patient discomfort and motion-induced artifacts. But, from the radiation exposure perspective, the ultra-fast acquisition isn’t any better. A key thing to emphasize here is that our system has better resolution in addition to better sensitivity which could enable the detection of small lesions in the heart and abnormal wall motion. This could potentially benefit the population in general as well as high-risk populations such as high-orbit astronauts.

Investigating the depth of interaction (DOI) effect in detail, building a low-cost design of an ellipsoid detector system, and integrating this with the latest electronic readout mechanism are the major future directions of this work.